\documentclass[10pt]{iopart}
\usepackage{graphicx}
\usepackage{subfigure}
\usepackage{iopams}
\usepackage[warning,math]{easyeqn}
\usepackage[thinlines]{easybmat}
\usepackage[square,sort&compress]{natbib}
\usepackage{array}
\newtheorem{conjecture}{Conjecture}[section]
\newcommand{\scs}{\scriptscriptstyle}

\newcommand{\xiB}{\mbox{\boldmath$\xi$}}

\def\a{\alpha}
\def\o{\omega}
\def\wz{\omega_{\scs 0}}

\def\b{\beta}

\def\e{\epsilon}

\def\<{\langle}
\def\>{\rangle}
\begin{document}
\title{Discrete breathers in protein structures} 
\author{Francesco Piazza$^1$,
        Yves--Henri Sanejouand$^2$}
\address{$^1$ Ecole Polytechnique F\'ed\'erale de Lausanne,
              Laboratoire de Biophysique Statistique, ITP--SB,  
              BSP-722, CH-1015 Lausanne, Switzerland \\
        $^2$ Ecole Normale Sup\'erieure,
             Laboratoire Joliot-Curie, CNRS-USR 3010,
             46 all\'ee d'Italie,   
             69364 Lyon Cedex 07, France}
\ead{Francesco.Piazza@epfl.ch}
\ead{Yves-Henri.Sanejouand@ens-lyon.fr}
%
%
%
%
%
%
\begin{abstract} 

Recently, using a numerical surface cooling approach, we have shown that 
highly energetic discrete breathers (DB) can form in the stiffest parts of
nonlinear network models of large protein structures.
In the present study, using an analytical approach, we extend our previous results to 
low-energy discrete breathers as well as to smaller proteins

We confirm and further scrutinize  the striking site selectiveness of energy 
localisation in the presence of spatial disorder. In particular,  we find that, as a sheer
consequence of disorder,  a non-zero energy gap for exciting a DB at a
given  site either exists or not. Remarkably, in the former case, the gaps arise as
result of the  impossibility of exciting small-amplitude  modes in the
first place. On the contrary, in the latter case, a small subset  of
linear edge modes act as accumulation points, whereby DBs can be
continued to arbitrary small energies, while unavoidably approaching one of
such normal  modes.   In particular,  the case of the edge mode seems peculiar, 
its dispersion relation being simple and little system-dependent.

Concerning the structure-dynamics relationship, we find that the regions of protein structures 
where DBs form easily (zero or small gaps) are unfailingly the most highly connected ones, also 
characterized by weak local clustering. Remarkably, a systematic analysis on a large database of enzyme structures
reveals that amino-acid residues involved
in enzymatic activity tend to be located in such regions. This finding reinforces
the idea that localised modes of nonlinear origin may play an important biological role, e.g. by 
providing a ready channel for energy storage  and/or contributing to lower energy barriers of chemical reactions.

\end{abstract} 


\vspace{2pc}

\submitto{\PB}

\maketitle

\section{Introduction}

Proteins are molecular machines whose functional motions 
are strongly related to, if not encoded within,
their three-dimensional structure~\cite{Best:2007qe,Torchia:2003oq,Anfinrud:1996eu}.
As a matter of fact, useful information on their functional motions can be 
obtained 
from the mere knowledge of their equilibrium structure, as solved for example through 
X-ray crystallography or NMR spectroscopy,
even at the level of the harmonic approximation of the 
system potential energy~\cite{Brooks:85,Marques:95,Perahia:95}. 
Remarkably, this
structure-dynamics-function relationship can be captured
even when substantial amounts of structural details are missing.
In particular, elastic network models (ENMs) of proteins~\cite{Tirion:1996mz,Bahar:97,Hinsen:98,NMA} 
have been used for describing quantitatively amino-acid fluctuations 
at room temperature~\cite{Tirion:1996mz}, 
often in very good agreement with isotropic~\cite{Bahar:97}, 
as well as with anisotropic measurements~\cite{Phillips:07,Maritan:02}.
They have also allowed to show
that a few low-frequency normal modes can often provide fair insight on
the large amplitude motions of proteins 
upon ligand binding~\cite{Tama:01,Delarue:02,Gerstein:02}, 
demonstrating the robust character of these collective 
motions~\cite{Ma:05b,Tama:06,Nicolay:06}. 
Taken together, such results highlight the important role of the peculiar equilibrium scaffolds
of proteins~\cite{Rueda:2007pd,NMA,Tama:01}, at the same time
providing a rationale for the coarse-graining of amino-acid assemblies.

Recently, the interest for problems potentially involving nonlinear effects in 
bio-molecules, such as 
localisation and storage of energy, has increased
in the community at the interface between physics and 
biology~\cite{breathprot,breath-macromol,kosevich-2007,Yakushevich:2002lr,Savin:2003fk,Hennig:2002fk,Hennig:2002lr}.
A hot case concerns enzymatic catalysis and, more specifically, the following question: 
how does an enzyme store and 
use the energy released at substrate binding or when a chemical bond is broken ? 
Noteworthy, it is known that this energy 
may be deployed on much longer time scales (microseconds to milliseconds) than those
characteristic of the energetic process,
and at very distant places with respect 
to the catalytic site of the enzyme~\cite{enzimes}, 
so that it is highly unlikely that a normal-mode assisted mechanism 
may prove enough for explaining the phenomenon.

Indeed, protein dynamics has been long known to be highly 
anharmonic~\cite{Levy:82,Go:95}, 
a property which is certainly important in order to understand energy storage and transfer 
as a consequence of ligand binding, chemical reaction, 
{\it etc}~\cite{Straub:00,YuX._jp026462b}. Among nonlinear effects, the possibility that localised 
vibrational modes  of nonlinear origin may play a role in biological processes has recently been put 
forward by many authors~\cite{breath-macromol}, based on engrossing experimental studies
reporting numerous subtleties of protein  dynamics, 
notably through infra-red  spectroscopy~\cite{YuX._jp026462b,nonlinspectr1,Xie:2000ys,Xie:2001fr}.
For example, the excitation of localised vibrations in $\alpha-$helices has been proposed 
to be a way for enzymes to store energy during catalysis~\cite{breathprot,solHbonds}.
Within this framework, energy transfer across  helices would occur predominantly by 
hoppings of localised  vibrations  along the chain resulting from nonlinear coupling 
of spatially overlapping localised modes  in resonance~\cite{nonlinspectr4,Leitner:01}.

Nonlinear excitations proposed to play an active role in protein functional dynamics 
include topological excitations such as solitons~\cite{dOvidio:2005qy,Scott:1992} as well
as discrete breathers (DB)~\cite{Archilla:2002lr,Aubry:01}.
The latter are nonlinear modes that  emerge in many contexts as a result of both nonlinearity 
and spatial discreteness~\cite{Flach:1998fj}. Their existence and stability properties are well understood in systems 
with translational invariance at zero temperature~\cite{Aubry:2006dq}, 
and are also intensively 
investigated for nonlinear dynamical systems in the presence
of a thermal environment~\cite{Rumpf:2007rr,Peyrard:1998nx,FLACH:1994wd,BURLAKOV:1990cr}.
However, not much is known regarding the subtle effects arising
from the interplay of spatial disorder and anharmonicity~\cite{DB:04,nonlin-disorder:01,Rasmussen:1999vn},
either in general or in the context of the functional dynamics of biological 
macro-molecules.

In a recent paper~\cite{Juanico:2007}, we have introduced the nonlinear network model (NNM), with the aim 
of studying the simplest model that would take into account  
both the topology of protein structures and the anharmonicity of interparticle
potentials. Building upon the many successes of ENMs~\cite{NMA},
in the NNM framework, a protein is mapped onto a coarse-grained network of oscillators, 
whose equilibrium positions reflect the spatial arrangement of amino-acids in the protein fold. 
Noteworthy, in the linear regime, at low temperature for instance, the dynamical
behaviour of NNMs and ENMs are identical. However, in the non-linear regime, 
the behaviour of NNMs becomes much more complex. For instance,
by applying the technique of surface cooling, we have demonstrated that discrete 
breathers form spontaneously at a small subset of specific sites in a given structure, 
invariably in the stiffest regions. 
Remarkably, an interpretation of this finding can be proposed in terms of 
enzyme functional dynamics since, 
by studying stiffness patterns across a large data set
of enzymes, we have shown that catalytic residues tend to sit in the most 
rigid portions of their structure. It is thus tempting to speculate that
enzymes may take advantage of the well known ability of discrete breathers to harvest and
retain for long periods of times amounts of energy much larger than what
is normally available at a given site at a given temperature,
in order to achieve their function, which requires  crossing energy barriers.

As it often happens, the preliminary findings reported in ref.~\cite{Juanico:2007} have 
raised numerous questions. In particular, they have revealed several striking features 
of DBs in NNMs of proteins which deserve further detailed analysis, like
the subtle, ubiquitous spatial modulation of their properties, such as 
their dispersion relation, connection with the edge normal modes, 
degree of localisation and presence of an energy gap in the excitation spectrum. 
On the other hand, the numerical simulations analyzed in ref.~\cite{Juanico:2007} allowed
us to study highly energetic DBs {\it only}, as a consequence of the lack of a simple and 
clear-cut criterium for recognizing low-energy ones.
Moreover, the protein surface cooling procedure allowed us to observe DBs in
large proteins {\it only}, like citrate synthase, 
a 2 $\times$ 371 amino-acids dimeric enzyme (PDB code 1ixe),
probably as a consequence
of a too large surface-volume ratio of the smaller proteins considered,
like HIV-1 protease, a 2 $\times$ 99 amino-acids dimeric enzyme (PDB code 1a30).

It is the purpose of this paper to present a first analytical study of discrete 
breathers in NNMs of proteins in order to study low-energy DBs, 
and to show that such DBs also exist in smaller proteins. Moreover,
this is an opportunity to start answering, {\it en passant},
a fair number of interesting questions raised by our previous findings.
The paper is organized as follows. 
In section~\ref{sec:1}, we discuss the main features of NNMs of proteins.
In section~\ref{sec:2}, we show how it is possible, in the case of such models,
to obtain approximate breather solutions for the equations of motion, 
based on a simple argument invoking a separation of timescales.
In particular, we discuss how we solved the crucial problem of determining a suitable 
initial guess, a problem common to the majority of analytical methods for 
finding DB orbits, which is here further 
non-trivially faceted by the presence of spatial disorder.
The localised solutions that we obtain, the spatial modulation of their basic properties 
and their connection with the static and dynamical
features of the protein folds are analyzed in depth in section~\ref{sec:3}.
Finally, in section~\ref{sec:4} we summarize our results and discuss the relevance of our
findings in the context of enzyme function, while, in section~\ref{sec:5},
work still under way is rapidly evoked.

%
\section{Nonlinear network models of proteins\label{sec:1}}
%
 
In this section, we describe the nonlinear network model (NNM). 
Like in elastic network models (ENMs), a given protein is modeled as an 
ensemble of $N$ fictitious particles occupying the equilibrium positions of the 
$\alpha$-carbons, as found in the experimental structure.
All particles have the same mass~\footnote{As our fictitious particles 
occupy the equilibrium 
positions of amino-acids, i.e. are located on the corresponding $\alpha$-carbons,
we will use the words particles and (amino-acid) residues interchangeably.}, 
which we set equal to the average amino-acid mass $M = 120$ Da, and
each particle interacts with its neighbours,
as specified by a cutoff distance $R_{c}$. More precisely, 
residues $i$ and $j$ interact if $|\bi{R}_{i}-\bi{R}_{j}| \leq R_{c}$,
where $\bi{R}_{i}$ denotes the position vector of the $i$-th residue in the 
equilibrium structure. 

Let $\bi{u}_{i} = \bi{r}_{i}-\bi{R}_{i}$ denote the displacement vector of the 
$i$-th residue, $\bi{r}_{i}$ being its instantaneous position. 
In the central force approximation, the inter-particle potential energies may be expanded
in power series as follows 
\begin{eqnarray}
\label{e:pot2}
U(\bi{u}_{i},\bi{u}_{j})   &=& \sum_{p=2}^4 
                                  \frac{k_{p}}{p} \, \left(r_{ij} - R_{ij} \right)^p + \mathcal{O}(u^5) \\
                           &=& \sum_{p=2}^4 
                                  \frac{k_{p}}{p} \, (|\bi{u}_{ij} + \bi{R}_{ij}| -  R_{ij})^p
                                  + \mathcal{O}(u^5)                                  
\end{eqnarray}
where $\bi{u}_{ij} = \bi{u}_{i}  - \bi{u}_{j}$ and
$r_{ij} = |\bi{r}_{ij}|Ê= |\bi{r}_{i}-\bi{r}_{j}|$,
$R_{ij} = |\bi{R}_{ij}|Ê= |\bi{R}_{i}-\bi{R}_{j}|$ are the inter-particle instantaneous and
equilibrium distances, respectively. The total potential energy can then be written as
\begin{equation}
\label{e:pottot} 
\mathcal{U}(\{ \bi{u} \}) = \sum_{i} \e_{i}
\end{equation} 
where we have introduced the site energies $\e_{i}$
\begin{equation}
\label{e:siteen}
\e_{i} = \frac{1}{2} \sum_{j \neq i} c_{ij} \, U(\bi{u}_{i},\bi{u}_{j}) 
\end{equation} 
specified through the contact matrix
\begin{equation}
\label{e:contmap}
c_{ij}= 
\left\{
\begin{array}{l l } 
\theta(R_{c} - R_{ij})  & i \neq j \\
0                       & i = j
\end{array}
\right .
\end{equation}
$\theta(x)$ denoting the Heaviside step function. 
Setting $k_{3}=k_{4}=0$ in Eq.~\eqref{e:pot2} amounts to building a network of Hookean
springs joining pairs of atoms separated by a distance smaller than $R_{c}$,
that is, an ENM~\cite{Tirion:1996mz,Bahar:97,NMA}.
Here, we wish to study the simplest model capturing the combined effects of spatial
disorder and nonlinearity. Hence, we restrict ourselves to symmetric potentials, by
setting $k_{3}=0$. This choice allows us to get rid of distinct
nonlinear features associated with asymmetric terms of the potentials, such as 
kinks and more complicated DC components of the localised modes~\cite{Flach:1998fj}, 
that are likely to interact with topological disorder in peculiar ways and, as such,
deserve special attention in their own right.

The equations of motion for the $m$-th residue then read
\begin{equation}
\label{e:totFm}
\fl \ddot{\bi{u}}_{m}  = \wz^2
                       \sum_{j \neq m} c_{mj} 
                       \left[
                      \frac{     (|\bi{u}_{jm} + \bi{R}_{jm}| -  R_{jm}) + 
                           \beta \, (|\bi{u}_{jm} + \bi{R}_{jm}| -  R_{jm})^3}
                           {|\bi{u}_{jm} + \bi{R}_{jm}|}
                       \right] \, (\bi{u}_{jm} + \bi{R}_{jm})                                       
\end{equation}
where we have introduced the natural frequency $\wz = \sqrt{k_{2}/M}$ and the 
parameter $\beta = k_{4}/k_{2}$.

We want now to look for solutions of the equations of motion in the form of localised, 
time-periodic modes with angular frequency lying above
the linear spectrum\footnote{We observe that, in principle, 
the connectivity matrix $c$ may be such that a sufficiently extended zone of forbidden 
frequencies in the linear spectrum might exist, in  which case localised solutions 
with their frequency located in such a gap might be possible.},
i.e. the spectrum of the 
Hessian matrix of the potential energy, as given by Eq.~\eqref{e:pottot}.
A sufficiently general ansatz has the form of a periodic sinusoidal oscillation modulated 
by a function of time that varies slowly on the timescale defined by the 
inverse frequency $\o^{-1}$,
\begin{equation}
\label{e:DBans}
\bi{u}_{m}(t) = A \, \xiB_{m}(t) \cos \omega t
\end{equation}
We assume that the envelope functions $\xiB_{m}(t) $ are bounded and such that
$\max \xiB_{m}(t) \simeq \mathcal{O}(1)$, so that the amplitude $A$ sets the 
physical scale for the oscillation amplitude. 

In order for a physically sensible solution with its largest displacement at a given site $k$
to exist, we must require $A \ll \min_{j} R_{jk}$,
that is the maximum vibration amplitude of the mode must be much smaller than 
the shortest bond between the $k$-th particle 
and its neighbours\footnote{To be more quantitative: 
the Lindemann criterium~\cite{Lindemann} for avoiding (local) melting in 
classical systems prescribes $A < A_{\rm melt} \simeq 0.17 \min_{j} R_{jk}$.}.
Therefore, we can substitute the ansatz~\eqref{e:DBans} into equations~\eqref{e:totFm} and 
expand each addendum in the sum  over $j$ in series of $A/R_{jm}$. After some lengthy 
but straightforward algebra, we get:
\begin{eqnarray}
\label{e:totFmexp}
\fl \left( \frac{\o}{\wz} \right) ^2  A \xiB_{m}  C_{\o} + 
\left( \frac{2 \o}{\wz^2} \right) A \dot{\xiB}_{m}  S_{\o} -
\frac{A \ddot{\xiB}_{m}}{\wz^2}   C_{\o} =  \\ \nonumber
                    \fl - \sum_{j \neq m} c_{mj} 
                      \Bigg\lbrace 
                        \bi{R}_{jm} 
                         \Bigg[ 
                           (\hat{\bi{R}}_{jm} \cdot \Delta \xiB_{jm}) \, \e_{jm} C_{\o} +
                           \left( 
                             \frac{\Delta \xi^2_{jm}}{2} - 
                             \frac{3 (\hat{\bi{R}}_{jm} \cdot \Delta \xiB_{jm})^2}{2} 
                           \right) \e^2_{jm} C_{\o}^2 
                           \\  \nonumber
                           \fl + \left( 
                                \frac{5 (\hat{\bi{R}}_{jm} \cdot \Delta \xiB_{jm})^3}{2} 
                             -  \frac{3 \Delta \xi^2_{jm} (\hat{\bi{R}}_{jm} \cdot \Delta \xiB_{jm})}{2} 
                             + \beta R^2_{jm} (\hat{\bi{R}}_{jm} \cdot \Delta \xiB_{jm})^3
                           \right) \e^3_{jm} C_{\o}^3 
                         \Bigg] 
                         \\  \nonumber 
                          \fl + R_{jm} \Delta \xiB_{jm}
                             \Bigg[ 
                               (\hat{\bi{R}}_{jm} \cdot \Delta \xiB_{jm}) \, \e^2_{jm} C_{\o}^2  +
                               \left( 
                                \frac{\Delta \xi^2_{jm}}{2} - 
                                \frac{3 (\hat{\bi{R}}_{jm} \cdot \Delta \xiB_{jm})^2}{2} 
                               \right) \e^3_{jm} C_{\o}^3 
                             \Bigg]            
                      \Bigg\rbrace + \mathcal{O}(\e^4)
\end{eqnarray}       
where $C_{\o}=\cos \o t$, $S_{\o}=\sin \o t$, $\e_{jm} = A/R_{jm}$ are the expansion parameters, 
$\Delta \xiB_{jm} = \xiB_{j} - \xiB_{m}$, the relative displacement 
patterns, and where we have introduced the unit distance vectors 
$\hat{\bi{R}}_{jm} = \bi{R}_{jm}/R_{jm}$.

Since we are assuming that the envelope functions are slowly varying during
one breather oscillation, we can multiply Eqs.~\eqref{e:totFmexp} by $\cos \o t$ and average over one
period $2\pi/\o$. By the same token, we neglect the second time derivatives of the envelope 
functions. By doing this, we finally obtain a nonlinear algebraic system of $3N$ equations 
whose $3N+1$ unknowns are the time-averaged envelope patterns and the breather frequency:
\begin{eqnarray}
\label{e:eqmotalg}
\fl \left( \frac{\o}{\wz} \right)^2  \xiB_{m} = 
                    - \sum_{j \neq m} c_{mj} 
                      \Bigg\lbrace 
                        \hat{\bi{R}}_{jm} (\hat{\bi{R}}_{jm} \cdot \Delta \xiB_{jm}) 
                            + \frac{3A^2}{8R^2_{jm}} \times \\ \nonumber
                           \Bigg[
                             \hat{\bi{R}}_{jm}
                               \Big(
                                   (5 + 2 \beta R^2_{jm}) (\hat{\bi{R}}_{jm} \cdot \Delta \xiB_{jm})^3
                                   - 3 \Delta \xi^2_{jm} (\hat{\bi{R}}_{jm} \cdot \Delta \xiB_{jm})
                                \Big) \\Ê\nonumber +
                        \Delta \xiB_{jm} 
                                \Big(
                                  \Delta \xi^2_{jm} - 3 (\hat{\bi{R}}_{jm} \cdot \Delta \xiB_{jm})^2
                                 \Big)
                              \Bigg]          
                      \Bigg\rbrace 
\end{eqnarray}           
We note that the system of equations~\eqref{e:eqmotalg} is only 
apparently underdetermined. In fact, 
one can normalize the displacement pattern  $\xiB$ by taking any of its $3N$ components 
as the reference unit length. In practice, we shall keep the DB amplitude
$|\xiB_{m}| = A_{\scs B}$  fixed   for a guess mode centered at site $m$, 
which means solving  for the $3N$ variables 
$\{\xiB_{1},\xiB_{2},\dots,\varphi_{m},\vartheta_{m}, \dots, \xiB_{N},\o_{\scs B}\}$ 
where $\varphi_{m}$ and $\vartheta_{m}$ are the  azimuthal and polar angles, respectively, 
of the vector $\xiB_{m}$ and where $\o_{\scs B}$ is the DB frequency.

Equations~\eqref{e:eqmotalg} constitute a generalization to arbitrary topology of
well known equations for approximate breather solutions in periodic lattices,
which have been shown to produce accurate results in that context~\cite{Sandusky:1992dp}.
We note that, in analogy to lattice systems, we would have obtained the same set of 
equations by applying the so-called rotating-wave approximation (RWA), which amounts to truncating 
the expansion of $\cos^3 \o t$ to the first harmonic~\cite{Flach:1998fj}. 
However, RWA does not set
restrictions on the time variation of the envelope functions. Hence, we prefer to follow 
the time-averaging approach since it reflects more correctly the physical situation.
For vanishing amplitude $A$, we recover the eigenvalue problem of the elastic network -- 
in the limit $A \rightarrow 0$,
Eqs.~\eqref{e:eqmotalg} give the normal modes (NM) and the linear
spectrum of the system. In expanded notation, the corresponding eigenproblem reads
\begin{equation}
\label{e:linspec}
\left( \frac{\o}{\wz} \right)^2 \xi^\a_{m} = -
\sum_{\beta=x,y,z} \sum_{j \neq m} \mathbb{K}^{\a \b}_{mj} \, \xi^\b_{j} 
\end{equation} 
where greek apexes indicate spatial directions.
The matrix $\mathbb{K}$ is the Hessian of the potential energy:
\begin{equation}
\label{e:Hessian}
\mathbb{K}^{\a \b}_{mj} = c_{mj} \hat{\bi{R}}^{\a}_{mj} \hat{\bi{R}}^\b_{mj} - \delta_{jm}
                                   \sum_{k \neq m} c_{mk} \hat{\bi{R}}^\alpha_{mk} \hat{\bi{R}}^\beta_{mk}                              
\end{equation}   
$ \delta_{jm}$ being the Kronecker symbol.

%
\subsection{Solving the equations: the initial guess problem\label{sec:2}}
%

We wish now to devise a general procedure for determining 
nonlinear localised solutions of Eqs.~\eqref{e:eqmotalg}. 
We note that, at variance with periodic media, 
sites in a spatially disordered network are not equivalent. 
Indeed, the connectivity, i.e. the number of neighbours, 
as specified by the cutoff distance $R_{c}$, 
varies from site to site and so do the directions of the outgoing bonds, 
so that each position has a unique local neighbourhood. 
Consequently, it is natural to expect that the existence of 
localised solutions and their properties will be subject to unknown positional constraints.
In fact, the breathers obtained with the cooling procedure 
turned out to be site-selective, 
revealing that only a small fraction of the available sites would {\em support}
a breather spontaneously self-exciting out of the energy drainage process at the surface.
In particular, DBs single out special regions of the structure,
namely, those that are both among the most connected 
and the most buried ones.
Moreover, we also found a high degree of site dependence in frequency-energy dispersion relations, 
suggesting that also distinctive low-energy properties, such as the occurrence and the nature 
of an energy threshold for localisation, might be spatially modulated.

\begin{figure}[t!]
\centering
\resizebox{80mm}{!}{\includegraphics[clip]{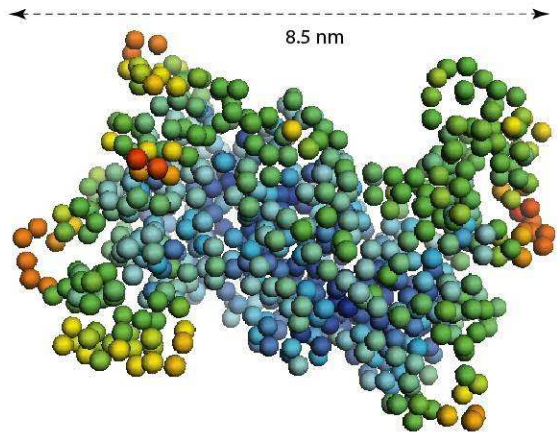}}
\caption{Structure of dimeric Citrate Synthase (PDB code 1IXE). 
Only the $\alpha$-carbons are shown, as spheres
in a color scale corresponding to the crystallographic B-factors,
from smaller (blue) to larger (yellow) fluctuations (color online).
\label{f:CITS}}
\end{figure}

In line with the above observations, we assume that
Eqs.~\eqref{e:eqmotalg} will converge to a breather solution centered at a given site $m$
provided $(i)$ the site $m$ {\em admits} a breather solution at all at the given 
point in  parameter space (the parameters of the model being $R_{c}$, $\beta$ and either
the DB frequency or its amplitude\footnote{By that we shall always mean the 
maximum of the displacement field $A |\xiB_{m}|$, $m=1, 2, \dots, N$, 
where $\{\xiB_{m}\}$ is the calculated time-average DB displacement pattern. }) and 
$(ii)$ the initial guess is close enough to a nonlinear periodic solution.
Let us stress that we are explicitly assuming that, as a consequence of the 
non-equivalence of the sites, due to topological disorder, 
DB families are also indexed by
the site at which they are localised. Such an assumption 
is strongly supported by our previous numerical results, since
all DBs obtained proved to be higly localised ones, noteworthy, more localised
than any of the harmonic modes~\cite{Juanico:2007}.
In practice, condition $(i)$ should be regarded as a criterium for convergence within 
a specified tolerance for the sum of residuals. 
Importantly, such a criterium must be 
restricted to a homogeneous, site-independent protocol for identifying the initial guess.
In other words, a protocol able to identify magnitudes and direction cosines at each site 
for the guess displacement field automatically, without arbitrariness, e.g. as what 
concerns the number of displaced particles, the directions of displacements {\em etc}.

An analysis of the mode patterns obtained in ref.~\cite{Juanico:2007} from a Principal Component 
Analysis of the system trajectories in the quasi-stationary state revealed 
that such DBs typically concentrate the largest part of their energy on 
a single residue, leaving only a fraction of the order $1/\mathcal{N}$  
on each of its  $\mathcal{N}$ neighbours. Furthermore, a closer inspection allowed
to show that the displacement vector of the central residue and those of its neighbours
seem to be oriented so as to maximize the associated distortion of the network. 
All other residues being virtually at rest for a typical DB eigenvector, this 
also correspond to maximizing the system potential energy~\eqref{e:pottot}.
The above observations suggest the following protocol for computing the
initial guess to be fed to system~\eqref{e:eqmotalg}. 

Suppose we wish to solve Eqs.~\eqref{e:eqmotalg} for 
a DB of given amplitude $A_{\scs B}$ at site $m$. The $3N$ unknowns are then $3N-1$ time-averaged 
displacement coordinates and the frequency $\omega_{\scs B}$. Alternatively, we could as well fix the frequency
and solve for the pattern and the amplitude.
We start with the network in  its equilibrium configuration.
The first step consists in drawing at random the 
azimuthal and polar angles of the displacement vector $\xiB_{m}$, with the constraint 
$|\xiB_{m}| = A_{\scs B}$, so as to uniformly sample the ensemble of all vectors 
centered at site $m$ with a fixed modulus $A_{\scs B}$.
Typically, looping a number $\mathcal{O}(10^3)$ of times, we get a
satisfactory convergence of the maximum strain direction at site $m$ with a relative error of 
$1-2$ \%. We then repeat the same operation sequentially at all the $\mathcal{N}(m)$ neighbouring sites,
in ascending order with respect to the bond distances $R_{jm}$ $(j=1, 2, \dots, \mathcal{N}(m))$, with the only 
difference that also the magnitude of the displacements is varied.
Following the numerical results of ref.~\cite{Juanico:2007},
we take as initial guess for the magnitudes 
$|\xiB_{j}| = A_{\scs B}/(\mathcal{N}(m)+1)$ for all $\mathcal{N}(m)$ neighbours.
The calculation proceeds site by site
in such a way that, when optimizing the displacement at site $j$, the {\em shells} closer 
to the central residue are kept fixed in the previously determined optimal configurations.
We coin this procedure the sequential-maximum-strain (SMS) method. 

\begin{figure}[t!]
\centering
\resizebox{110mm}{!}{\includegraphics[clip]{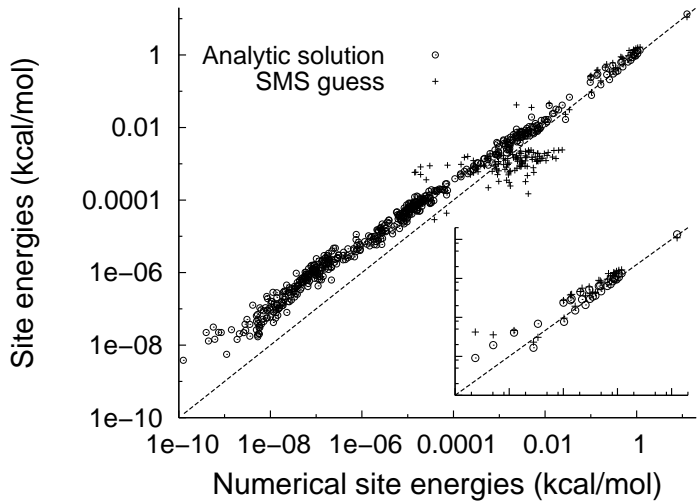}}
\caption{Discrete breathers at site THR 208 A of Citrate Synthase. 
Comparison of the site energies of the analytic breather solution
and of the SMS pattern with a numerical DB
obtained through the cooling process with amplitude $A_{\scs B} = 0.929$ \AA; see ref.~\cite{Juanico:2007}. 
The dashed line is the plane bisector.
The inset displays the site energies of all 32 neighbours
of the central residue.
Parameters of the NNM are $R_{c} = 10$ \AA, $\beta = 1$ \AA$^{-2}$.\label{f:DB206energies}}
\end{figure}

An illustration of how the SMS protocol performs is given in Fig.~\ref{f:DB206energies},
where we show a scatter plot of the site potential energies of a numerical solution from 
ref.~\cite{Juanico:2007} versus the corresponding SMS guess vector as well as the 
analytic solution for the case of the site associated to Threonine 208, in 
monomer A of Citrate Synthase (see Fig.~\ref{f:CITS}).
Note the wide separation of energies (logarithmic scale)
between the central site and the neighbouring
sites, as a consequence of the localised character of highly energetic DBs. 
As a further quantitative confirmation that
SMS-based analytical breathers reproduce well those obtained during our 
previous cooling simulations, 
we note that the frequency of the analytic
solution (Eqs.~\eqref{e:eqmotalg}), also shown in Fig.~\ref{f:DB206energies}, 
is $\o_{\scs B} = 115.3$ cm$^{-1}$, as 
compared to that determined by the numerics, namely, 
$\o_{\scs B} = 114 \pm 1$ cm$^{-1}$.
However, despite the excellent predictive power of the SMS-fed protocol, 
the lowest site energies of the numeric solution tend to be smaller with respect to the 
analytic solution. Since the breather is exponentially localised, these correspond to 
sites located far from the DB centre, which are likely to be 
close to, if not on, the protein surface. 
Although the effect is a small-amplitude one, it seems to be an interesting 
consequence  of the surface cooling protocol itself, whereby breathers would appear further {\em squeezed}
into the interior of the structure so as to minimize its amplitude in the surface 
regions where the protein is in contact with the zero-temperature bath.

\section{Results: analytical discrete breather solutions in NNMs of proteins\label{sec:3}}

In the previous section we have discussed the general
features of NNMs of proteins, along with a strategy for
obtaining approximate analytic solutions. Since we are
interested in computing localised modes whose properties
are expected to be site-dependent, we have devised a
homogenous protocol for obtaining a well-defined guess
for a DB at a given site. To be more explicit, the SMS
method clarifies ambiguous points related to computing
an initial guess for the DB pattern such as the number
of displaced residues, the magnitude and directions of
their initial displacements and so forth, the algorithm automatically
taking care of these choices in compliance with a general requirement. 
Furthermore, we have shown that the pattern computed through
the SMS procedure is remarkably close to the numerical
DB obtained by surface cooling. In this section, we
shall discuss the properties of the localised vibrations calculated
from equations~\eqref{e:eqmotalg}, starting from SMS guesses. 

\subsection{Insight from the cooling simulations: high-energy DBs}

\begin{figure}[t!]
\centering
\begin{tabular}{c c}
\resizebox{65mm}{!}{\includegraphics[clip]{THR208-ALA209-ALA212_WE.eps}} &
\resizebox{65mm}{!}{\includegraphics[clip]{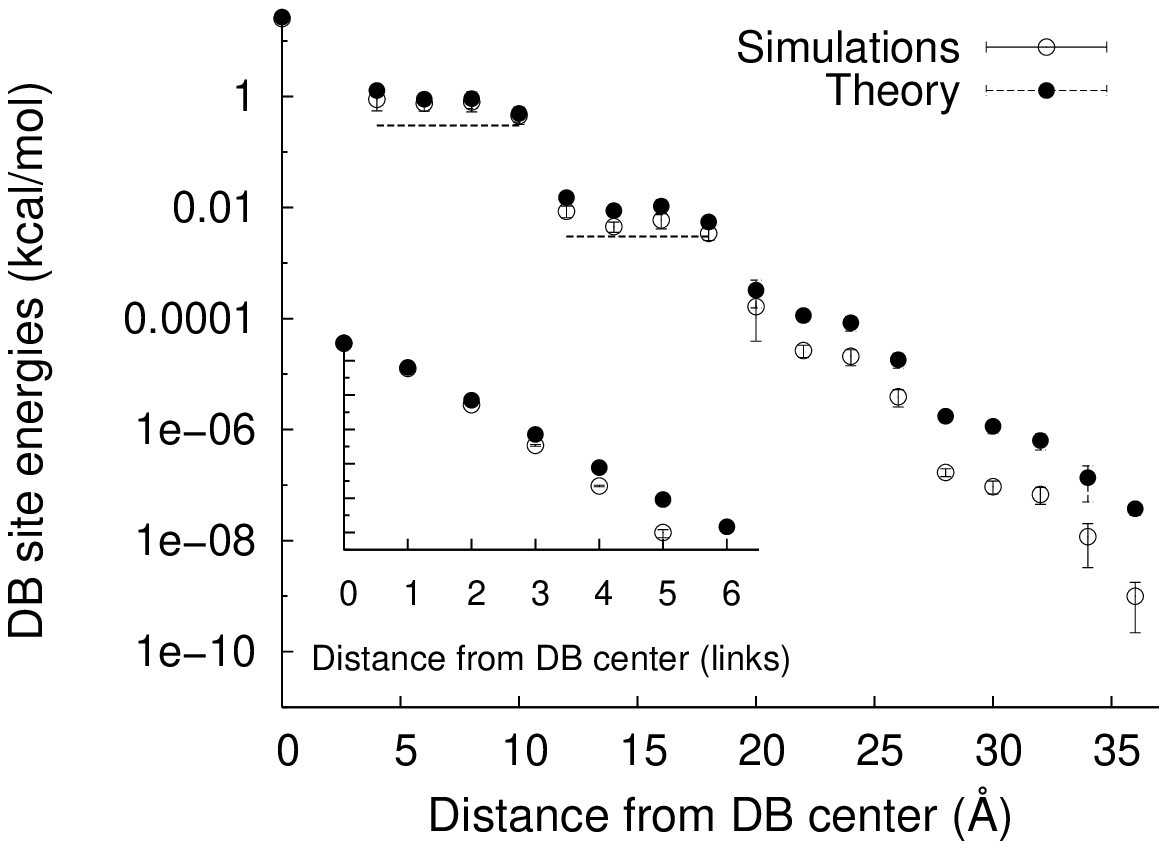}} \\
\resizebox{65mm}{!}{\includegraphics[clip]{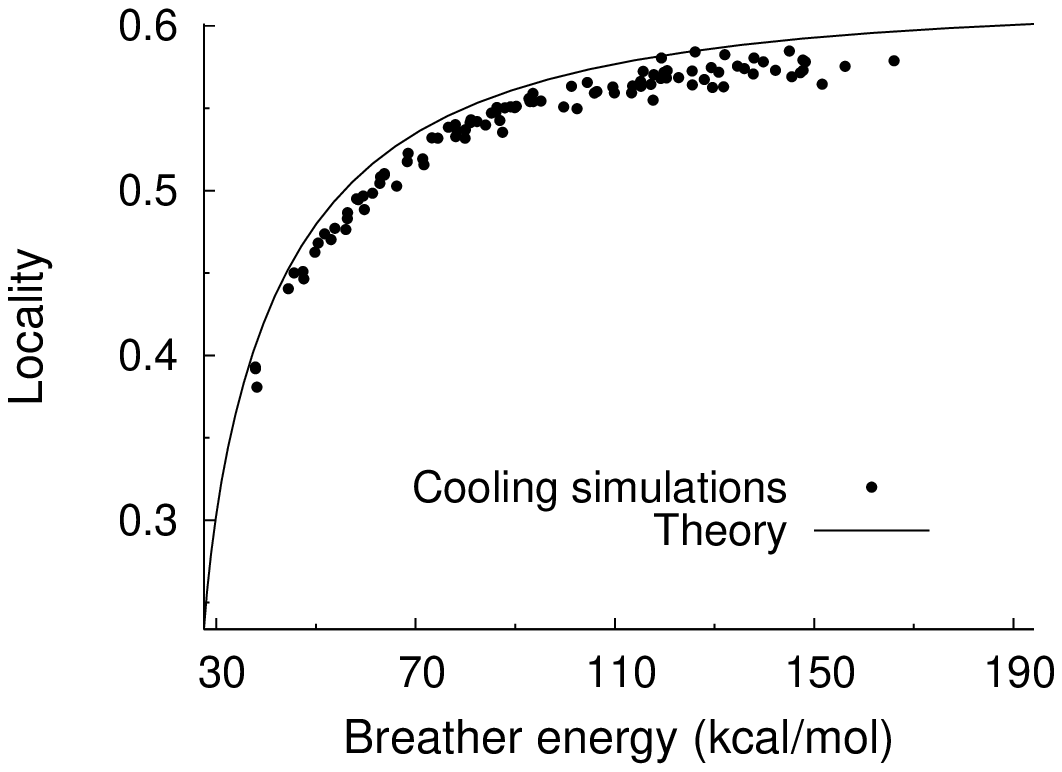}} &
\resizebox{65mm}{!}{\includegraphics[clip]{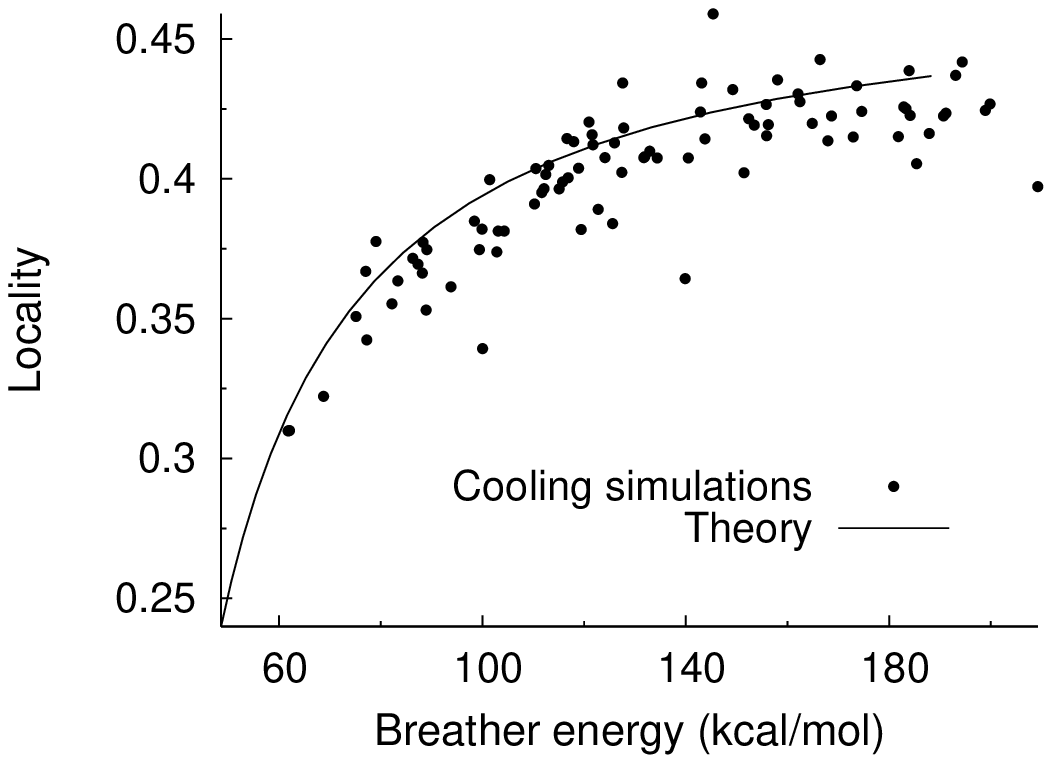}} 
\end{tabular}
\caption{Comparison  between analytical and numerical 
DBs, as obtained in cooling experiments with Citrate Synthase.
Upper panels: numerical (circles) and analytical (solid lines) dispersion relations for DBs self-localising 
on monomer A (left) and site energies versus distance from the central site (right)
for the DB at THR 208 A with amplitude $A_{\scs B} = 0.929$ \AA.
The inset shows the same energies versus average distance,
 expressed now in terms of number of links from the central site.
The $y$-axis units are the same as in the main plot.
Lower panels: locality measures as defined in Eq.~\eqref{e:Locyh} 
for DBs localised at THR 208 A (left) and THR 192 B (right).
Note that in the left panel, for example, the theory is only about 2 \% 
off the numerical results. Parameters of the NNM are $R_{c}=10$ \AA$, \beta = 1$ \AA$^{-2}$.\label{f:WE}}
\end{figure}

We have seen that SMS guesses approximate well
the patterns of the self-localising DBs obtained
by surface cooling in ref.~\cite{Juanico:2007}. 
It turns out that the analytical solutions 
that we compute starting from an SMS guess preserve, and even improve,
such agreement throughout the whole energy range spanned by the
numerical solutions. This point is illustrated in the case of Citrate Synthase 
in Figs.~\ref{f:WE}. The dispersion relations for the four DBs 
with the largest probabilities of occurrence in monomer A, as observed
in cooling  simulations (see table~\ref{t:DBcoolstat}),
are extremely well reproduced by our analytical approach. The spatial patterns are well reproduced too.
They can be measured through the locality index
\begin{equation}
\label{e:Locyh}
L = \frac{\sum_{i\alpha} \xi_{i\alpha}^4}{[\sum_{i\alpha} \xi_{i\alpha}^2]^2}
\end{equation}
where  $\xi_{i\alpha}$ is the $\alpha$ $(x, y, z)$ coordinate of the $i$-th atom
in the given displacement field $\xiB$.
So, our theoretical calculations confirm the strong site-to-site
variability of the DB dispersion 
relation, already spotlighted in ref.~\cite{Juanico:2007}.

As a further instructive comparison between the analytical solutions computed
from SMS guesses and those obtained through cooling simulations, 
Fig.~\ref{f:WE} also shows how 
site energies vary as a function of the distance from the central site of 
the DB (upper right panel). In particular, it shows that local 
energies decrease exponentially for sites farther and farther away
from the central site, as expected for a discrete breather.
However, a closer inspection of the curves reveals 
that site energies, at least those in the first coordination shells,
are sub-organized in plateaus. 
In other words, 
there are relatively ample (about 5 \AA) intervals of
distances from the central site where sites 
share more or less the same local energies.
This behaviour suggests that it is the number of links (bonds) between 
two sites rather than their distance that matters.
Indeed, a plot of site energies as a function 
of their separation from the central site, expressed in units of {\em links}, i.e. the 
distance in the sense of graph theory, 
confirms the validity of this conjecture (see inset of Fig.~\ref{f:WE}).
Here, the graph is the one defined 
by the connectivity matrix $c_{ij}$
and the distance between two nodes (the sites) is the smallest number
of links that have to be followed in order to go from one node to the other. 
As a matter of fact, by measuring distances in such units, 
the exponential decay of site energies is recovered.

The numerical breathers analyzed in Ref~\cite{Juanico:2007}
correspond to simulation runs (the majority) where a single object gathered 
nearly all the system energy. Moreover, 
as a criterium for selecting mere DBs, we explicitly required 
that the energy collected be larger than the initial energy per particle 
in the thermalized state. In other words, we selected and studied high-energy 
DB's {\it only}.  Hence, the comparison illustrated in Figs.~\ref{f:WE}
shows that the SMS scheme is surely a good procedure for computing guess modes 
for analytical DBs of relatively high energy. As shown hereafter,
the situation gets more complex when we turn to the
low-energy portions of the dispersion relation. 

\begin{table}
\caption{Statistics of DB self-localisation in dimeric Citrate Synthase,
as found in our previous cooling simulations;
see ref.~\cite{Juanico:2007}. The probability of appearance $p$ is the fraction of
localisation events out of a pool of 500 cases.
Note that the slight asymetry of citrate synthase has significant 
effects on the statistics of events in the two monomers.
\label{t:DBcoolstat}}
\begin{indented}
\item[]\begin{tabular}{@{}c c r c c r}
\br &\multicolumn{2}{l}{\bf Monomer A}&\multicolumn{1}{c}{}&\multicolumn{2}{l}{\bf Monomer B}\\
\cline{2-6}
Rank & Residue & $p$ (\%)  & & Residue & $p$ (\%)  \\
\mr
 1 &  THR 208	& 20.4   & &  THR 192  &   17.8    \\ 
 2 &  ALA 209	& 11.1   & &  ALA 196  &   12.5    \\   
 3 &  ALA 196	&  4.5   & &  THR 208  &    9.1    \\ 
 4 &  ALA 212	&  2.5   & &  ALA 209  &    5.3    \\ 
\br
\end{tabular}
\end{indented}
\end{table}

\subsection{Low-energy breathers: does an energy gap exist?}
%

In the context of regular lattices, interesting predictions have been formulated as to 
the presence of an energy gap in the excitation spectrum of discrete breathers, 
i.e. a finite energy threshold that has to be overcome in order to create a DB.
It has been shown that, depending on the spatial dimension and type of nonlinearity in the inter-particle 
potentials, a finite threshold may, or may not, exist 
(for systems of infinite size)~\cite{Flach:1997qy,Kastner:2004fk}.
As shown hereafter, in the context of spatially disordered systems, 
such a question has to be
formulated not only in terms of {\em existence} of a gap, but also 
in terms of its {\em nature}.  

In general, for Hamiltonian systems, discrete breathers occur in 
one-parameter families, indexed e.g. by their energy $E_{\scs B}$, 
their frequency $\o_{\scs B}$ or alternatively by their amplitude $A_{\scs B}$ 
(in the language of ansatz~\eqref{e:DBans}).
In practice, in the presence of 
hard-type non-linearities, when the
DB frequency is lowered toward the linear band edge frequency
$\o_{\scs E}$, the breather amplitude may or may not approach zero. 

In an ordered medium, if DBs
of arbitrary small amplitudes exist at all, they can only appear
in the limit $\o_{\scs B} \rightarrow \o_{\scs E}$, as a direct consequence
of the constraint of non-resonance with any of
the linear modes, $\omega_{\scs E}$ being the edge frequency. 
In this case, a detuning exponent
$z$ can be introduced~\cite{Flach:1997qy} and standard perturbation
theory used to determine both the exponent and the coefficient in the relation
\begin{equation}
\label{e:detuning}
| \o_{\scs B} - \o_{\scs E}| = c A_{\scs B}^{z}
\end{equation}
In general, it is possible to speculate that, if DBs of arbitrary
low amplitude exist in a discrete system,
then they emerge from linear edge modes by means
of a tangent bifurcation~\cite{Kastner:2004fk}. 
Moreover, this hypothesis can be
used to calculate explicitly the energy at which such a bifurcation occurs.
For example, in the case of periodic lattices, it has been shown that 
the bifurcation energy either vanishes or is asymptotically finite 
in the limit of large system size, depending
on the spatial dimension and on the nature of
the potentials~\cite{Flach:1997qy,Kastner:2004fk}.
Hence the conclusion that breather excitation spectra may, or may not, display
a finite energy gap, depending on the nature of the studied system.
It is important to stress that in an ordered system the bifurcation of the nonlinear 
edge mode marks a symmetry breaking, the emerging DB mode being exponentially localised,
while the edge linear mode has an extended pattern.
The question is, then, how this picture changes when spatial disorder enters the game.

In the context of weakly-coupled chains of oscillators with on-site
disordered potentials, it has been shown that low-amplitude breathers
exist, approaching linear eigenvectors for vanishing amplitudes. Remarkably, the 
latter patterns are localised  
due to disorder and, therefore, there is no symmetry breaking and the DBs approach
the linear modes {\em continuously}. Moreover, there are DBs 
originating not only from the edge mode, but also from modes at frequencies below the
band edge~\cite{Kopidakis:1999fk,Kopidakis:2000lr}. These DBs have been 
coined {\em intra-band breathers} and their frequencies have been shown to be 
dense within the gaps separating the frequencies of two consecutive linear modes.
Under the general hypothesis that such features do not depend on the type
of disorder, so long as (at least) the edge modes are localised, we may
conjecture that a similar picture would be observed in the presence of spatial disorder.
In topologically disordered systems, such as protein
structures, sites are not equivalent, and edge modes are
intrinsically localised, different modes having their
largest displacement in different regions of the structure.
Hence, we may expect that different families of 
DBs may exist, localised at different sites and approaching different 
edge normal modes for vanishing amplitudes.
In particular, we may expect that both the DB frequency and 
the corresponding pattern smoothly approach
asymptotically the values of the corresponding linear modes with no symmetry breaking at all.
In any case, such DBs would not show an energy gap in their spectrum.
However, linear modes are characterized by a variable degree of localisation, only 
a small fraction of them being strongly localised.
In the case of Citrate Synthase, for example, Eq.~\eqref{e:Locyh}
predicts a localisation parameter greater than the average value plus
two standard deviations for a mere 4 \%  of the modes.
Therefore, also in accordance with Refs.~\cite{Kopidakis:1999fk,Kopidakis:2000lr}, we 
can guess that the sites on which localised normal modes are centered will host zero-gap DBs,
continuously approaching the corresponding linear patterns as their amplitude is reduced toward
zero, but will it be the case for a generic site ?

In order to answer the above questions and test our conjectures, we shall proceed as
follows. First, we look for localised solutions centered at a given site,
 and parametrized by the amplitude in the sense of 
ansatz~\eqref{e:DBans}\footnote{In principle, in order to solve Eqs.~\eqref{e:eqmotalg} 
for a guess centered
at a given site, we could either fix the parameter $A$ and
solve for the DB pattern $\{\xiB\}$ and frequency $\o_{\scs B}$, or fix the
latter quantity and determine the pattern and amplitude parameter. For exploratory purposes, 
fixing A appears the more natural choice also in view of the SMS strategy for computing the 
initial guess, whereby an SMS pattern is found at a given
value of $A$. However, we have also checked that the same zeroes of Eqs.~\eqref{e:eqmotalg}  would be found for 
identical choices of parameters through the alternative strategy.}.
We fix $A_{\scs B} = A_{0}$ at the site of choice and use the SMS pattern of amplitude $A_{0}$ as initial guess.
Then, we progressively decrease the value of $A_{\scs B}$ in small steps. If the sum
of residuals of the optimal solution found for Eqs.~\eqref{e:eqmotalg} 
is lower than a specified small tolerance (close to machine
precision), we record a DB solution and calculate its physical properties. 
Then such solution is used as guess for calculating the DB for the following value of $A_{\scs B}$. 
If the chosen amplitudes are sufficiently 
closely spaced, this algorithm enables us to follow a given DB family centered 
at the site of choice~\cite{Kastner:2004fk}.  Besides the DB (potential) energy $E_{\scs B}$
and frequency $\o_{\scs B}$, we also calculate the localisation parameter, 
as defined in Eq.~\eqref{e:Locyh}.

\begin{figure}[t!]
\centering
\begin{tabular}{c c}
\resizebox{110mm}{!}{\includegraphics[clip]{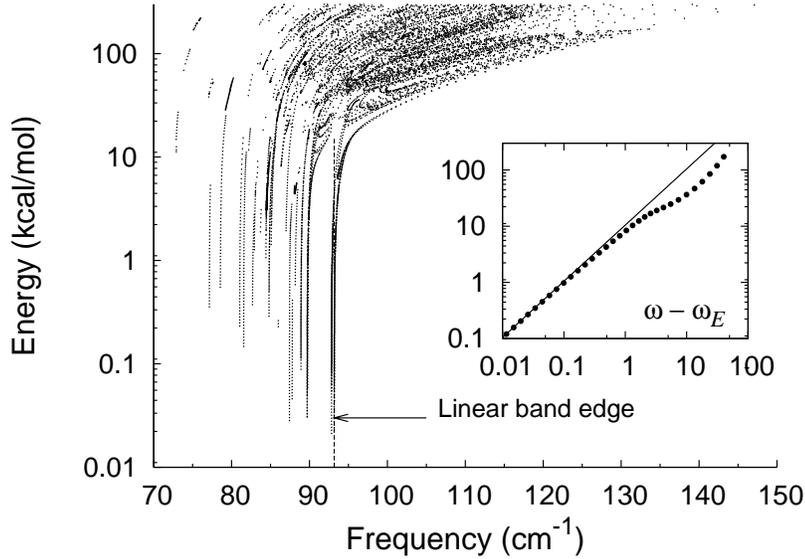}}
\end{tabular}
\caption{Dispersion relations for analytical DBs at all sites in HIV-1-protease (PDB 1A30). 
For the sake of clarity, only 
energies below 300 kcal/mol are shown. The inset shows a plot of energy versus frequency 
detuning from the edge normal mode, for the DB centered at site ILE 85A, the site with 
the largest displacement in the edge normal mode.
The solid line in the inset is the first-order perturbative prediction obtained 
through the Lindstedt-Poincar\'e method.
Parameters of the NNM are $R_{c}=10$ \AA$, \beta = 1$ \AA$^{-2}$.\label{f:WE.all1A30}}
\end{figure}

The results of the above-described calculations performed for all sites of a 
small dimeric enzyme (HIV-1 protease) are shown in 
Fig.~\ref{f:WE.all1A30}. The plot shows the dispersion relation for all DBs found, that is all points 
in the parameter space where a solution of system~\eqref{e:eqmotalg} could be found 
(to machine precision). 
The figure shows that intra-band breathers can exist within the framework of 
protein NNMs, i.e. localised modes of nonlinear origin with 
frequencies lying within the gaps of the linear spectrum. More precisely, we see that 
a limited number of linear frequencies act as {\em accumulation points} 
for the frequencies of DBs, which approach
asymptotically the corresponding normal mode in the limit of zero amplitude. The inset of 
Fig.~\ref{f:WE.all1A30} provides a clear-cut illustration of this phenomenon,
in the case of the DB approaching 
the edge mode. In real space, the DB pattern also approaches continuously the NM pattern.
We repeated similar calculations for different proteins, obtaining analogous results.
The rationale behind such observations can be resumed in the following two conjectures.

\begin{conjecture} 
Let us consider the site the most involved  in a given normal mode, 
i.e. the one whose displacement is the largest in the NM pattern (hereafter simply the {\em NM site}).
If a discrete breather centered at a given site $m$
can be found  at arbitrary small values of  its amplitude $A_{\scs B}$, its pattern will asymptotically 
tend to $\xiB^{[m]}$, that is the normal mode for which site $m$ is the NM site. Correspondingly, its frequency
will also approach the corresponding linear frequency $\omega^{[m]}$ in the limit $A_{\scs B} \rightarrow 0$.
If the linear frequency $\omega^{[m]}$ lies below the edge, the breather surely exists in the 
frequency interval $[\omega^{[m]},\omega^{[m+1]}]$, and may or may not exist for frequencies 
above $\omega^{[m+1]}$.
\end{conjecture}
 
As a matter of fact, there are sites that are {\em NM sites} of more than one
normal mode. In this case, starting from one of such NM sites, we find that DBs asymptotically approach the
normal mode with the highest frequency. 

The above conjecture provides a solid interpretative framework for 
what we shall call zero-gap breathers,
i.e. DBs that may be excited at arbitrary small energies.
The opposite inference, however,  is not always true, which means that not all DBs centered at a
given NM site approach the corresponding normal mode in the low-amplitude limit. There may simply be 
NO low-amplitude limit at a given site. In that case, at a given value of the amplitude, the DB solution
shifts from the selected site (that is the site we started from at amplitude $A_{0}$) to another 
site. Note that, in general, the destination site needs not necessarily be one of the neighbours of the original site.
In the following, we will refer to phenomena of the like as {\em jumps}. 
This brings us to formulate our second conjecture.

\begin{figure}[t!]
\centering
\begin{tabular}{c c}
\resizebox{65mm}{!}{\includegraphics[clip]{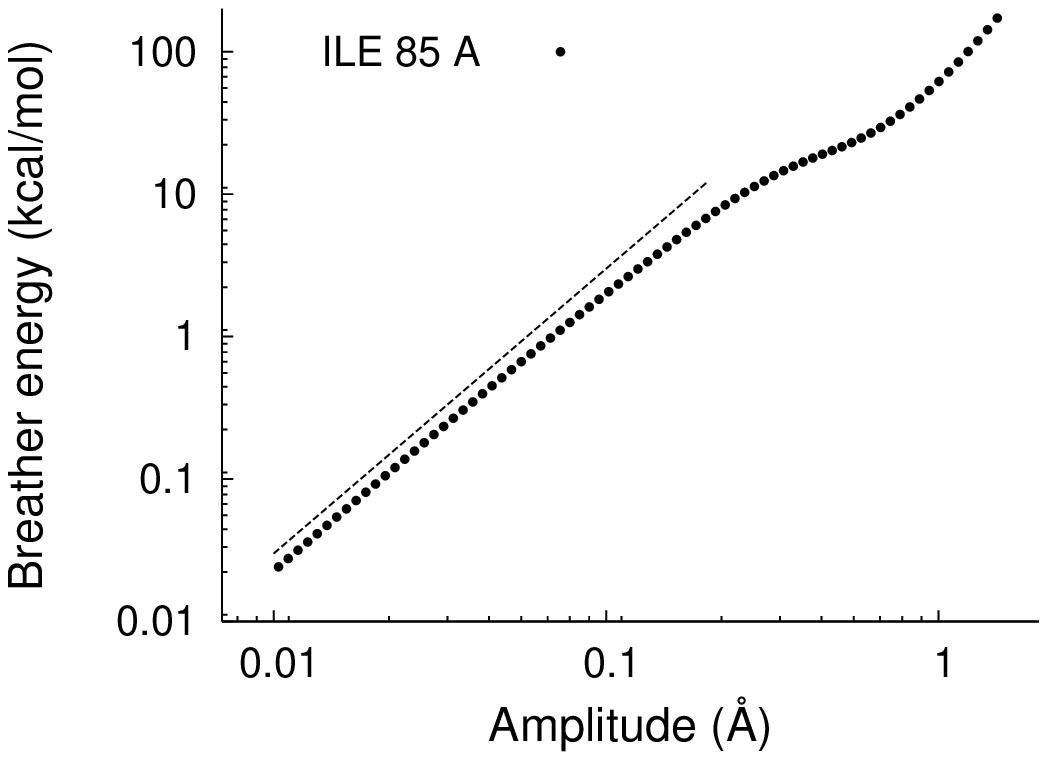}} &
\resizebox{65mm}{!}{\includegraphics[clip]{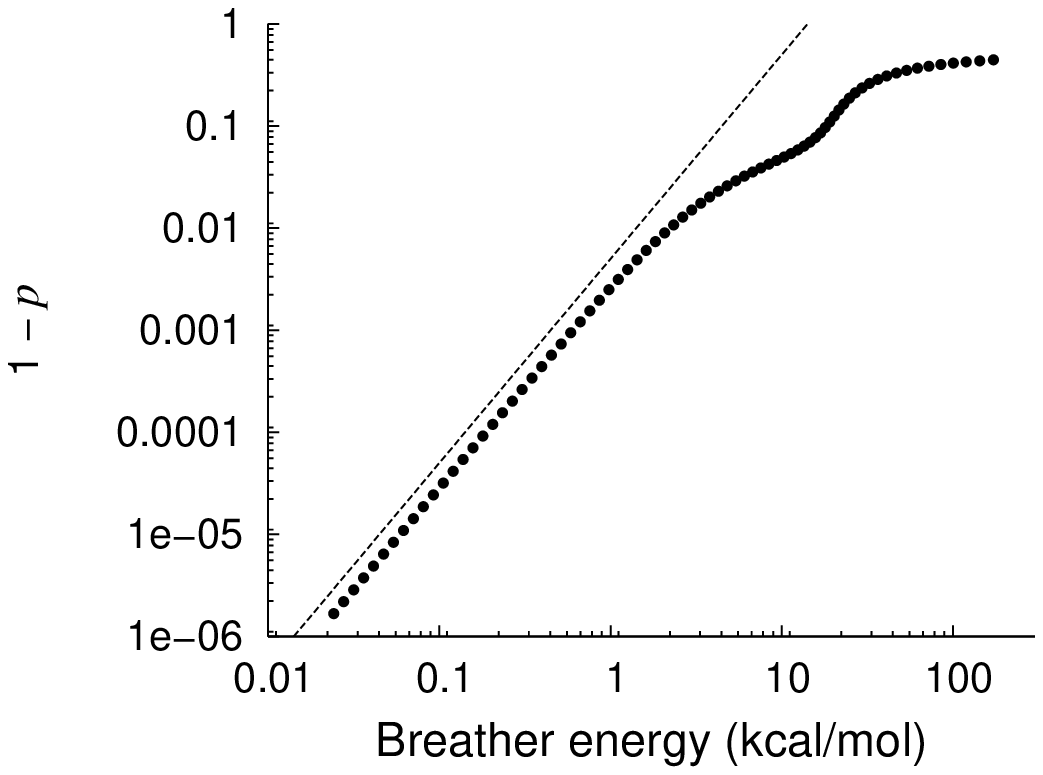}} \\
\resizebox{65mm}{!}{\includegraphics[clip]{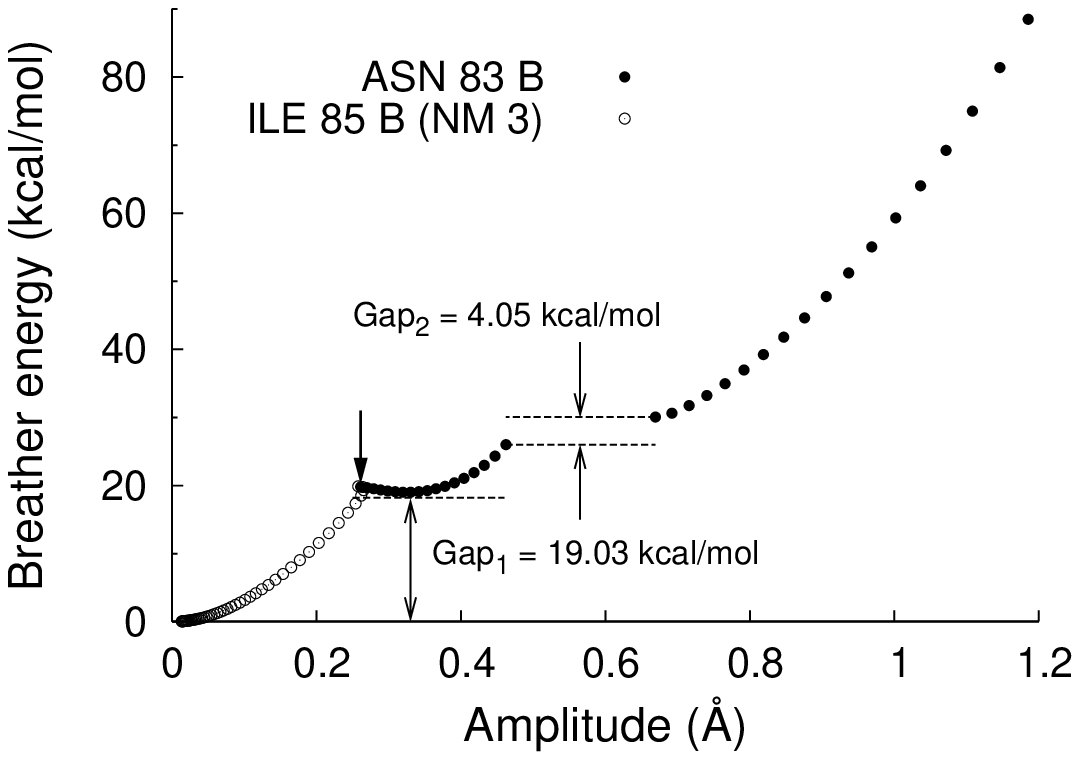}} &
\resizebox{65mm}{!}{\includegraphics[clip]{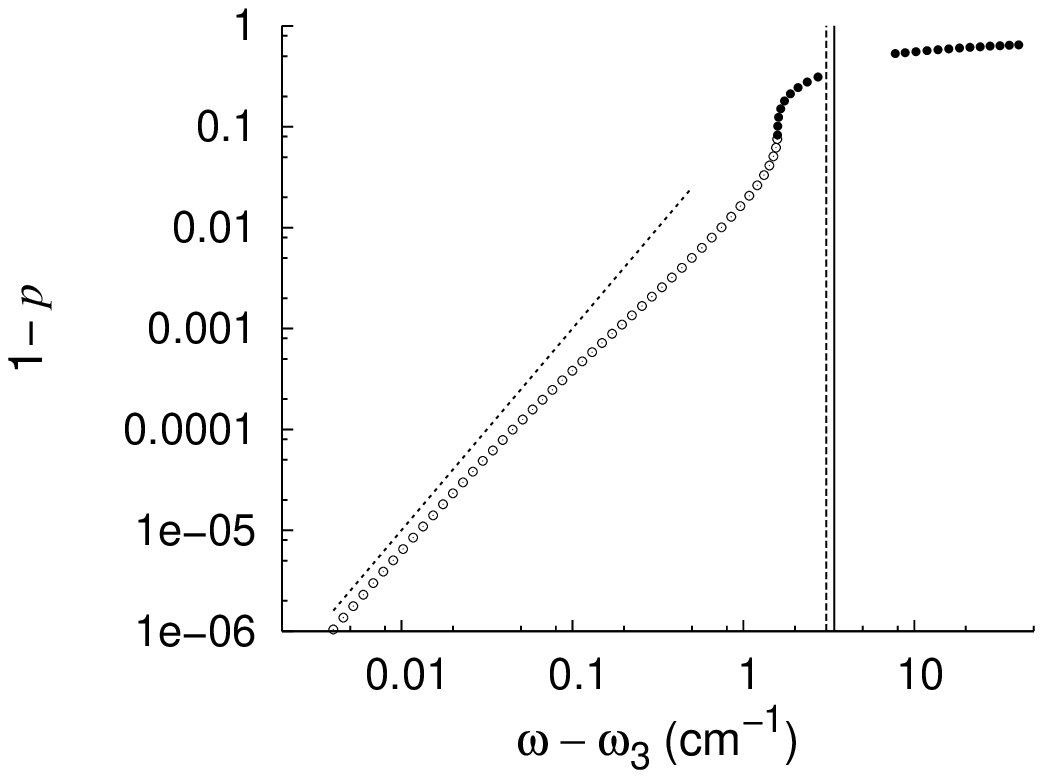}}
\end{tabular}
\caption{Examples of DB properties, in the case of HIV-1 protease (PDB 1A30). 
Left panels: plots of DB energy versus amplitude. Right panels:
(complementary) projection of the DB pattern on the 
corresponding asymptotic normal mode as a function of the DB energy (1$^{\rm st}$ mode, upper plot)
and of the frequency detuning from the mode (3$^{\rm rd}$ mode, lower plot). 
The dashed line in the upper left plot is a power law with detuning exponent $z=2$.
That in the upper right graph is a power law such that $1-p \propto E_{\scs B}^2$.
The lower graphs portrait a case where a jump occurs, from the original site ASN 83B (filled circles)
to the new site ILE 85B (empty circles).
In the right graph, the dotted line is a power law such that $1-p \propto (\omega - \omega_{\scs 3})^2$,
while the two vertical lines mark the edge mode and the second mode of frequency $\omega_{2}$.
Parameters of the NNM are $R_{c}=10$ \AA, $\beta = 1$ \AA$^{-2}$.\label{f:DBs-85-182}}
\end{figure}

\begin{conjecture} 
Let us suppose that a value of the amplitude at a given site $m$ has been reached where a jump has 
occurred to another site $n$, i.e. the largest amplitude in the DB patterns has shifted
from site $m$ to $n$. If, after the whole structure has been scrutinized, 
low-amplitude breathers are never recovered at site $m$ as the destination site 
of jumps observed from other sites during the whole analysis (covering all sites in the protein),
the DBs centered at site $m$ feature an energy gap in the spectrum, i.e. they exist
but for energies higher than a finite threshold. In this case, an energy gap exists as a direct 
consequence of the impossibility of exciting small-amplitude breathers in the first place.
\end{conjecture}

Note that, in the case of zero-gap breathers, the possibility still exists that an energy gap may open up 
at higher energies.  In those cases,  however, despite being centered at the same site, a change of 
symmetry occurs in the patterns
of the DBs lying above and below the gap. In this case, we shall speak of two different families of breathers.

The above considerations are illustrated in Fig.~\ref{f:DBs-85-182}, for the case of
the HIV-1 protease dimer. The upper panels refer
to the breather at the NM site of the edge normal mode. No jump is observed, and the DB
approaches continuously the edge normal mode with a detuning exponent  
$z=2$\footnote{At small energies, $E_{\scs B} \propto \omega - \omega_{\scs E} \propto A_{\scs B}^z$.}.
At variance with other properties of DBs in NNMs, 
we find that all DBs that exist up to arbitrary small amplitudes
are characterized by the same detuning exponent $z=2$.
Thus, the value of $z$ seems to be determined 
by the choice of the force field, and not by the topology of the structure. 
On the other hand, the normalized scalar product $p$ between 
the DB and the linear mode pattern approaches unity asymptotically 
when $E_{\scs B} \rightarrow 0$
following a power law, such that $1-p \propto E_{\scs B}^2$ (see upper right panel). 
Note that this DB is able of harvesting a  substantial amount of energy even at relatively small amplitudes, 
thus providing an easily accessible nonlinear energy storage channel. For example, typical 
energies involved in ATP hydrolysis, the most common fueling mechanism for molecular
machineries, are of the order of $10$ kcal/mol. A DB at site ILE 85A
would raise a similar amount of energy at a mere 0.2 \AA $\,$ of amplitude.

The lower plots in Fig.~\ref{f:DBs-85-182} show a case where jumps
are observed. Starting at large amplitudes with an SMS guess centered at site ASN 83B, 
a DB solution can be followed until an amplitude $A = 0.669$ \AA. Beyond that point, no 
DB solution is found, until the amplitude reaches the value $A=0.462$ \AA. At that point a
DB solution is recovered, centered at the same site. 
However, the lower right panel shows 
that a change in symmetry has occurred, as revealed by taking the normalized projection onto the 
3$^{\rm rd}$ mode from the edge as reference. Overall, 
there is a 4 kcal/mol-wide interval where no DB seems to exist.
Lowering the amplitude further, a new jump is observed at $A=0.261$ \AA.  At this stage,
the DB jumps to a neighbouring site (ILE 85B, which is about 7 \AA $ \ $ away) and from this point it 
approaches asymptotically   the corresponding normal mode (i.e. the one for which  ILE 85B is the NM site).
Therefore, we conclude that the DB centered at ASN 83B has a second energy gap (which marks 
the lower bound of its dispersion relation) of 19.03 kcal/mol, as explicitly 
reported in the lower left graph.
As it is clear from the lower right panel in Fig.~\ref{f:DBs-85-182}, the solutions that we find in the
frequency range below $\omega_{\scs 2}$
are intra-band breathers. 
Starting from the SMS guess for ASN 83B at high energies, we are able to continue the solution 
until a frequency slightly above the band edge. Decreasing the amplitude further beyond this point,
we no longer find any DB solution, until the intra-band DB centered at site ILE 85B 
is recovered, just below the second linear mode. 
This solution can then be further continued to approach asymptotically the third mode.

\begin{figure}[t!]
\centering
\begin{tabular}{c c}
\resizebox{65mm}{!}{\includegraphics[clip]{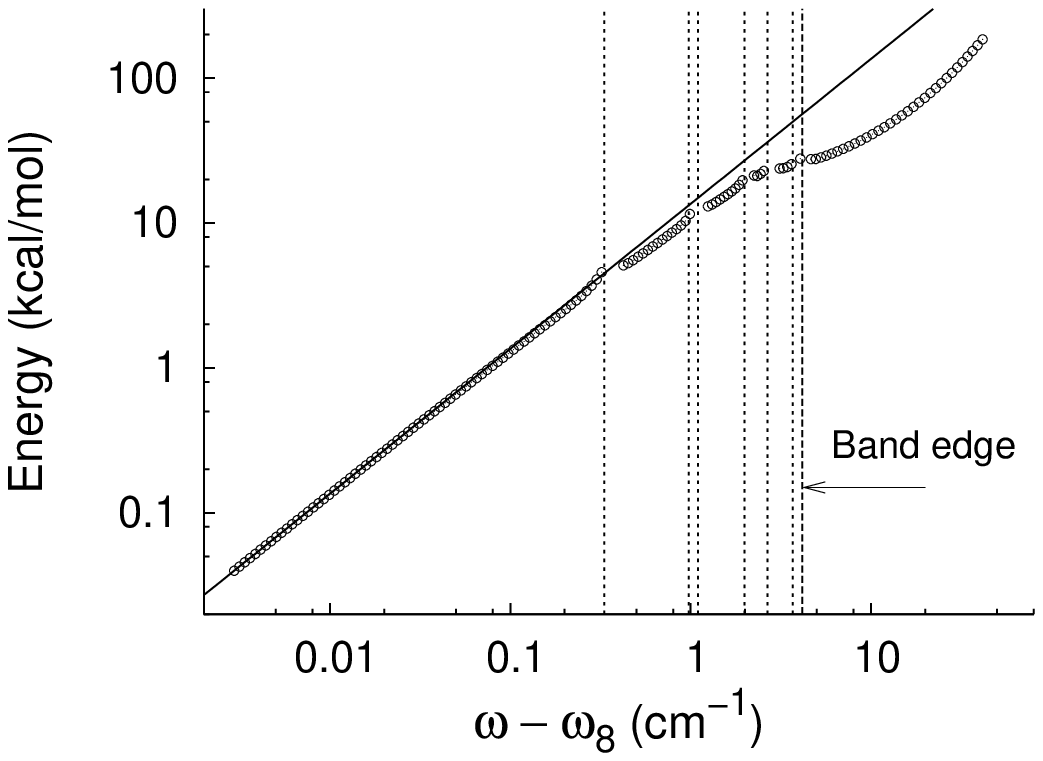}} & 
\resizebox{65mm}{!}{\includegraphics[clip]{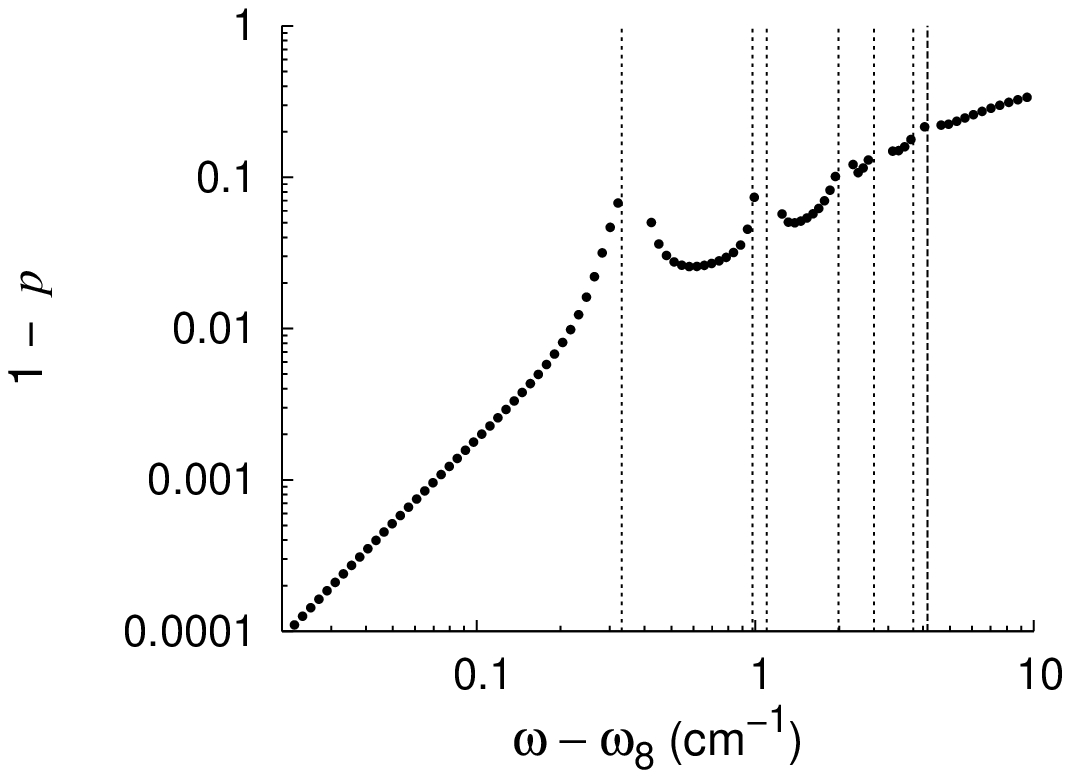}} \\
\resizebox{65mm}{!}{\includegraphics[clip]{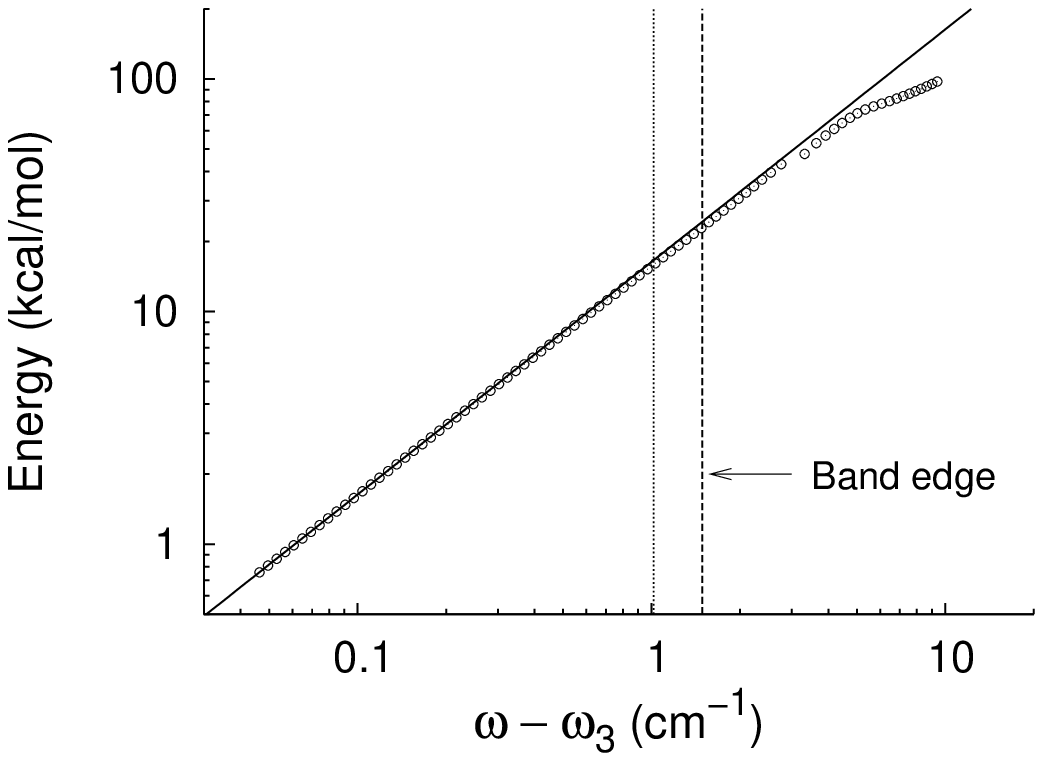}} & 
\resizebox{65mm}{!}{\includegraphics[clip]{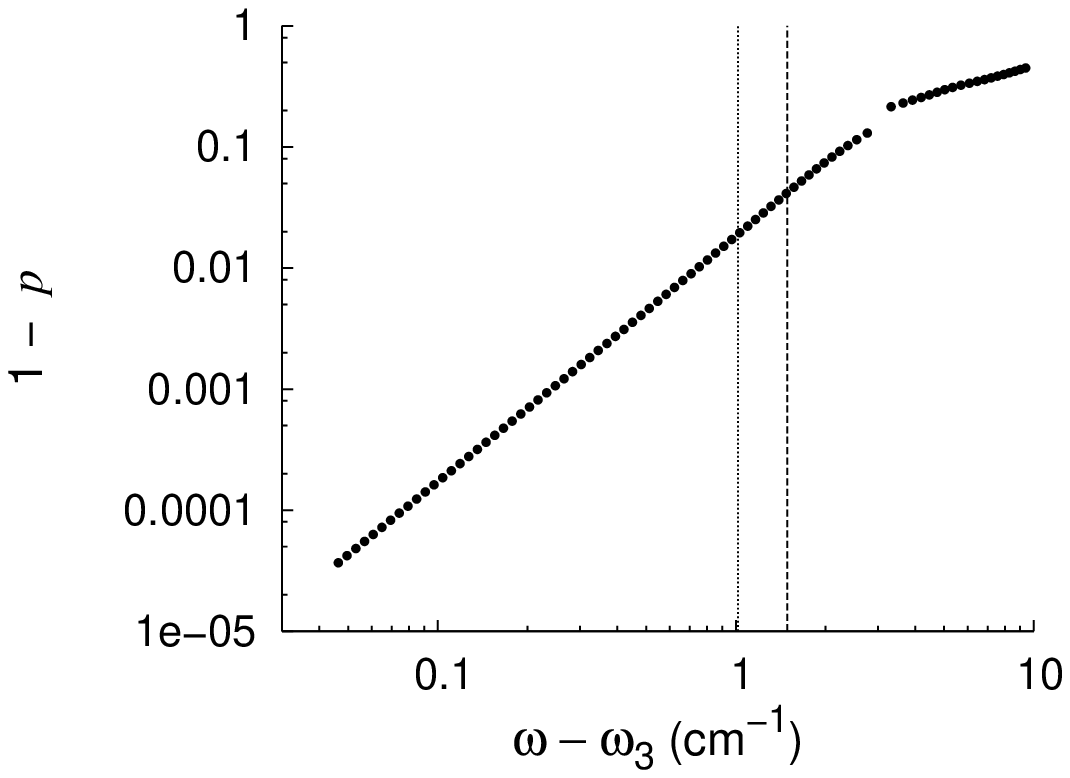}}
\end{tabular}
\caption{Energy (left) and normalized projection on the corresponding  asymptotic normal mode (right)
versus frequency detuning for the DBs at site GLY 255 A (upper panels) and 
ALA 196 B (lower panels) of Citrate Synthase,
approaching the 8$^{\rm th}$ and 3$^{\rm rd}$ highest normal modes, respectively. 
The solid lines  in the left panels are the result of first-order 
Lindstedt-Poincar\`e perturbation theory. The vertical 
lines mark the highest normal modes.\label{f:EW212-568}}
\end{figure}

\subsection{Zero-gap intra-band breathers: crossing harmonic levels}

In general, zero-gap DBs approach a given normal mode in the low-energy limit, their frequency
tending asymptotically to the corresponding eigenfrequency. However, we found that
the behaviour of such solutions may vary to a substantial extent from mode to mode, that 
is, from site to site.
The only exception is the case of DBs tending to the edge normal mode, which show a great regularity.
In fact, we find that the same behaviour illustrated in Fig.~\ref{f:DBs-85-182} is reproduced for all DBs 
approaching the highest frequency mode in all analyzed structures. 
In view of their robustness, that is, the reproducibility of their characteristics 
across different structures, such breathers may well happen to play a special role in NNMs of proteins.
Concerning DBs originating from modes lying below the edge, we find a much more faceted situation.
A deep and thorough understanding of the regularities, if any, displayed by such DBs in relation with 
their spatial environment and with the edge normal mode patterns and frequencies,
surely an instructive study, 
extends beyond the purpose of the present communication. Here, we limit ourselves to 
two other demonstrative examples, found when studying the case of dimeric
Citrate Synthase. 

If the {\em target} normal mode corresponds to a frequency  below the edge, the DB encounters a number 
of harmonic levels as its energy decreases. In such cases, the DB can be lost at one, 
several, or even at all such crossings, as we proceed from large amplitudes and a solution correspondingly 
recovered in the frequency interval between one mode
and the adjacent one. As a matter of fact, we speak of
intra-band DBs, as these are solutions that exist {\em only} within inter-mode gaps. 
Rigorously, a DB cannot exist at the exact frequency of a normal mode. Hence, adjacent NMs
may be thought as boundaries separating different DB families. However, it has been 
shown in another context that breather frequencies are dense within successive harmonic 
levels, which thus act as accumulation points~\cite{Kopidakis:1999fk,Kopidakis:2000lr}.

In general, discontinuities seem to occur in the majority of cases as we continue a DB solution through the
harmonic spectrum. One such example is reported in the upper graphs of Fig.~\ref{f:EW212-568}. 
Here, the DB eventually approaches the 
8$^{\rm th}$ highest frequency mode and the crossings with the seven harmonic 
levels above can be guessed simply by looking at the dispersion relation. 
In fact, an inspection of the projection of the DB on the corresponding normal
mode pattern allows one to spotlight the crossings,
and suggests that these are accompanied by 
changes in the symmetry of the solution. 
Thus, we may speculate that such intra-band
DBs belong to different families, each of them existing only within the gap between two given frequencies.
Remarkably, however, this is not the only behaviour of breathers as they encounter linear frequencies.
In fact, some DBs do not seem to be perturbed at all 
by crossing one or more linear modes. For example, the dispersion relation reported
in the lower left graph of Fig.~\ref{f:EW212-568} shows no discontinuities as the DB crosses the first 
(edge) and second highest modes, as it approaches the third highest mode. The projection of the DB mode on 
its asymptotic  pattern confirms that the breather is able to travel across two harmonic levels
virtually undisturbed, with no appreciable perturbation of its pattern. 

The observation of an instability when a DB crosses an normal mode close to the band edge 
can be rationalized by examining the spatial overlap between the two patterns. Since both modes are exponentially localised, 
if the NM is localised far from the DB core, an instability will not be detectable numerically within machine precision. 
Conversely, if the DB is centered in a region where also the edge mode is confined, our algorithm will signal the 
instability. In this sense, our results point to the the existence of {\em spatial selection rules}, that 
govern the range of existence/stability of a DB located at a given site.
It should be stressed that a more rigorous analysis based on numerical continuation algorithms such as the 
one employed in refs.~\cite{Kopidakis:1999fk,Kopidakis:2000lr} would allow a closer inspection of 
the bifurcations arising at the unstable level crossings, through e.g. a Floquet analysis of the DB 
stability. Work in this direction in the framework of NNMs of proteins is currently under way.

\begin{figure}[t!]
\centering
\begin{tabular}{c c}
\resizebox{110mm}{!}{\includegraphics[clip]{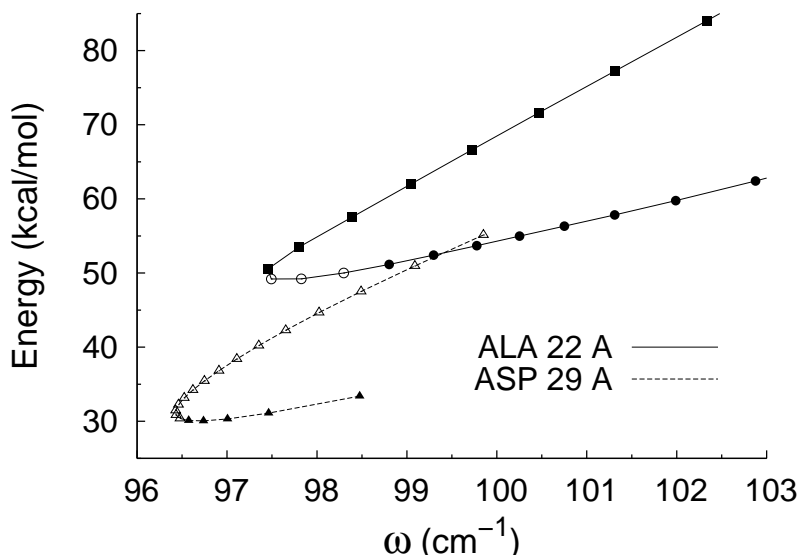}}
\end{tabular}
\caption{Discrete breathers in HIV-1-protease (PDB 1A30). Magnification of the dispersion relations 
of two DBs highlighting the occurrence of the bifurcation tongues. The change of symbols
along the curves marks a change in the DB symmetry, i.e. the shift of the DB centre 
on a different site.
Parameters of the NNM are $R_{c}=10$ \AA$, \beta = 1$ \AA$^{-2}$.\label{f:EW-022-029}}
\end{figure}

Finally, we wish to draw attention to another interesting feature displayed by our DB solutions
that was also reported in ref.~\cite{Kopidakis:1999fk}, namely the presence of {\em tongues}
in the dispersion relations of some DBs (see Fig.~\ref{f:EW-022-029}). These solutions display
a distinctive turning at a specific point in the energy-frequency plane, marking a finite
energy gap. Such tongues signal the occurrence in the vicinity of the turning point 
of a change of symmetry in the DB pattern, accompanied by a jump of the DB centre to a neighbouring 
site. In analogy to what reported in the case of the systems analyzed in ref.~\cite{Kopidakis:1999fk},
we conjecture that such jumps also mark a change in stability of the DBs through 
the appearance of a tangential bifurcation.
However, our approximate method do not allow us to perform an accurate linear stability analysis of 
our solutions, and hence we shall defer further statements on that matter to our subsequent studies.
We wish to stress, however,  that the possibility of following the (putatively) unstable branches is here
afforded by our choice to fix the amplitude at a given site and solve for the DB pattern 
along with its frequency. Had we followed one such breather by fixing its frequency,
we would have lost the solution at the turning point.

\section{Discussion~\label{sec:4}}

The complete analysis of HIV-1 protease reveals that 75.6 \% of sites are characterized  by a non-zero energy gap, while
14.7 \% have a vanishing gap and 9.7 \% of the sites do not support any DB at all, in the sense
that large amplitude solutions are found to shift to other sites.
Of course, we speak here of the lowest energy gap among all possible {\em holes}
that a single DB dispersion relation may display (see again the case of ASN 83B in 
Fig.~\ref{f:DBs-85-182}), that is, the particular non-zero value of the DB energy 
below which no solution centered at the studied site is ever recovered. 
An analysis of a selection (10 \%) of sites in Citrate Synthase yields similar figures: 80 \% of non-zero gap DBs, 17 \% of zero-gap
DBs and 3 \% of sites that do not seem to allow for any DB at all. 
It should be stressed again that the presence of sites that do not allow for DBs might
be regarded as a consequence of the upper bound chosen for the amplitude,
namely: $A_{\scs 0} = 1.5$ \AA. It is likely
that DBs could be recovered, at some of those sites, using larger values
for the amplitude. However, the energy gap of such DBs is expected to
be quite large, well over values that are expected to be sustainable for
chemically bound molecules (i.e. 100-200 kcal/mole).
In this sense, the presence in a given structure of regions where DBs
are not found has to be regarded as a direct consequence of the requirement 
of chemical meaningfulness.

\begin{figure}[t!]
\centering
\begin{tabular}{c c}
\resizebox{65mm}{!}{\includegraphics[clip]{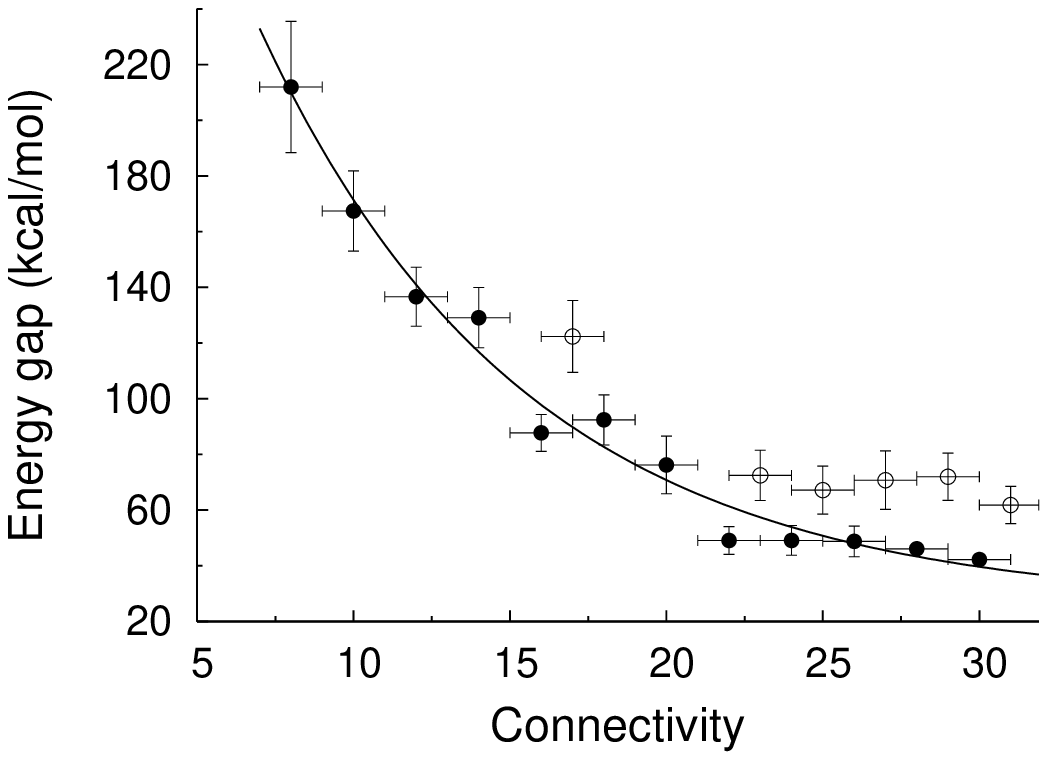}} &
\resizebox{65mm}{!}{\includegraphics[clip]{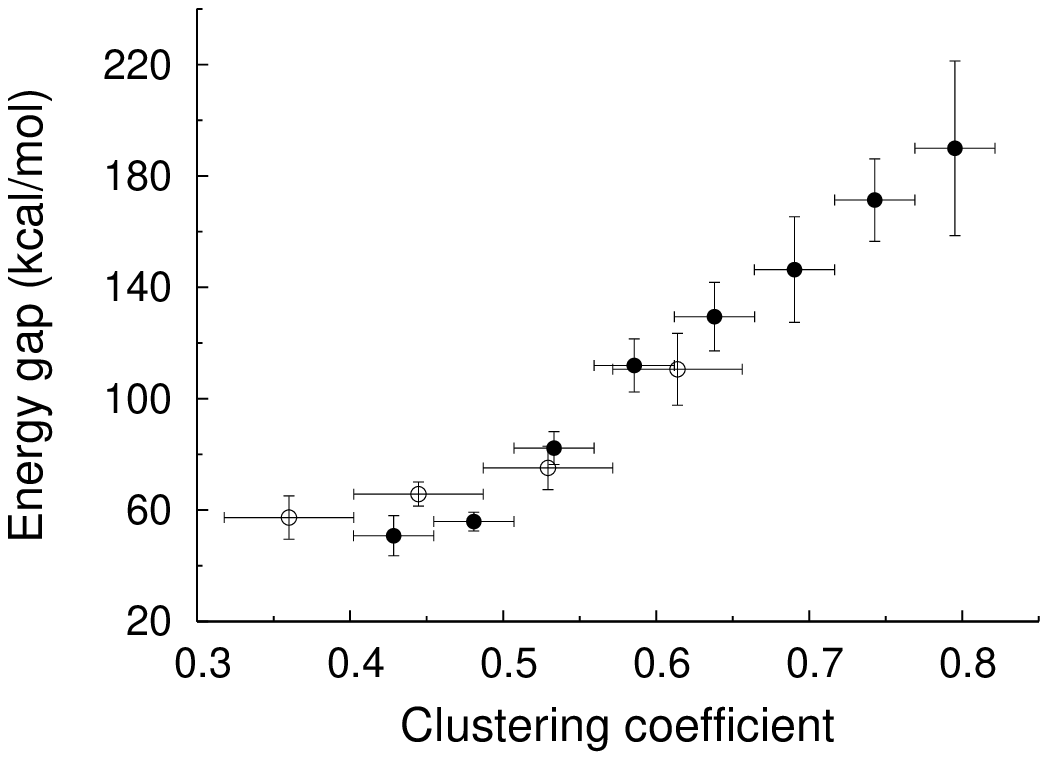}} \\
\resizebox{65mm}{!}{\includegraphics[clip]{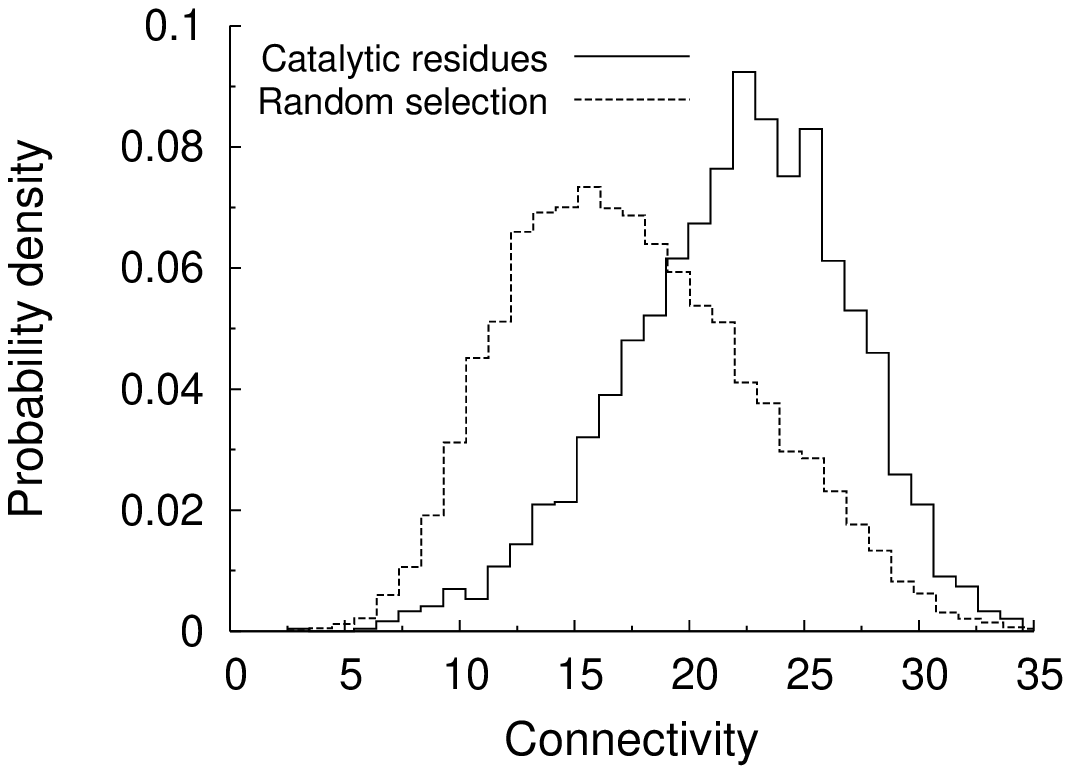}} &
\resizebox{65mm}{!}{\includegraphics[clip]{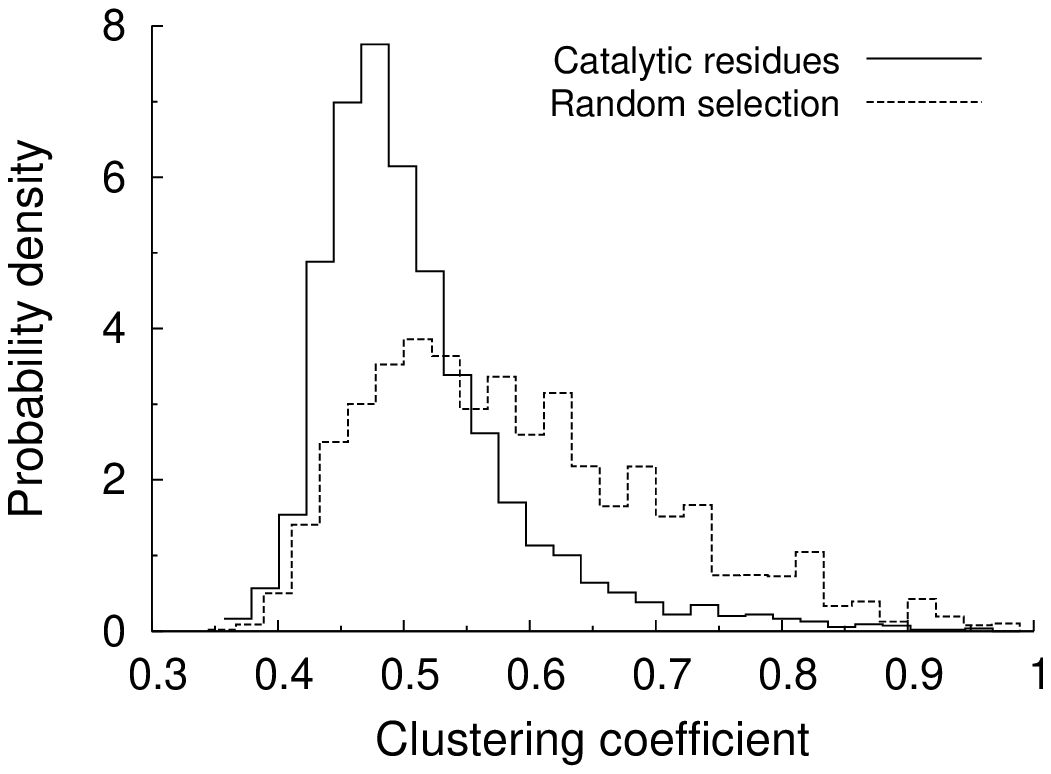}}
\end{tabular}
\caption{Upper panels. DB energy gaps versus connectivity (left) and clustering coefficient (right)
in HIV-1-protease (PDB 1A30, filled circles) and Citrate Synthase (PDB 1IXE, empty circles). 
The solid line in the left panel is just to guide the eye.
The error bars are the statistical errors associated with the binning of energy, connectivity
and clustering axes.
Parameters of the NNM are $R_{c}=10$ \AA$, \beta = 1$ \AA$^{-2}$.
Lower panels. Distribution of connectivities (left) and clustering coefficients (right) 
of amino-acid residues 
involved in enzymatic activity, compared to amino acids of same chemical type,
 randomly chosen within the same set of enzyme structures.\label{f:GAPS.cC3}}
\end{figure}

Since DBs with a finite energy gap seem to be the largest majority, we may ask what is 
the variability of energy gaps at all such sites, for a given structure, and investigate
the relationship between gap magnitudes and the structural properties of 
local neighbourhoods.
In general, gaps turn out to cover a rather wide energy range, 
between $\sim$40 and $\sim$200 kcal/mol. 
The lower bound is consistent with the lowest-energy DBs that we were able to excite 
by surface cooling~\cite{Juanico:2007}. On the other end, 
as shown in the upper plots of Fig.~\ref{f:GAPS.cC3}, the largest gaps found 
are clearly associated
with regions that are both poorly connected and
surrounded by a highly connected neighbourhood.
The latter fact can be quantified for a given site $i$ by the fraction of its 
neighbours that are also interacting, namely its {\em clustering coefficient},
\begin{equation}
\label{e:C3}
C_{3}(i) = \frac{2}{\mathcal{N}(i)[\mathcal{N}(i)-1]} \sum_{j>k} c_{ij} c_{jk} c_{ki}
\end{equation}
where $\mathcal{N}(i)$ is the number of neighbours of the $i$-th residue (that is, its connectivity)
and $c_{ij}$ is the $(ij)$ element of the connectivity matrix.
On the other side, the most easily excitable DBs are those centered at highly connected sites
with poorly connected neighbourhoods. 
It should be noted that the two above-illustrated 
correlations are not entirely independent, as there is a general average tendency for highly 
connected sites to be surrounded by less connected neighbours. 
However, as illustrated in Fig.~\ref{f:GAPS.cC3},
this fact provides a non-trivial reading frame for the gap-structure relationship,

Intuitively, a non-zero gap for DB formation may arise, as the amplitude is lowered,
as a consequence of the interplay beween two 
competing mechanisms. First, obviously, the DB energy has to
decrease when its amplitude is decreasing. 
However, as its energy drops, the DB becomes
less localised. Consequently, the connectivities of the most distant regions from the DB center
become crucial, as it may happen that the {\em integral} over highly connected regions
of many exponentially small terms results in a finite energy value,
even for vanishingly small DB amplitudes\footnote{Of course this argument is rigorous only in 
the limit of infinite system size.
However, it can still be employed to {\em rank} putative energy gaps in different situations 
depending on the corresponding connectivity patterns in the DB tail regions.}.
Incidentally, this is the mechanism by which the spatial dimension 
enters the game in determining the presence of a finite threshold 
in a periodic medium~\cite{Kastner:2004fk}, since in such cases the spatial dimension
determines the degree of connectivity of each site.
In the present case, we have seen that a zero energy gap marks DB solutions 
that exist for arbitrary small amplitudes, their energy also approaching zero 
as the amplitude vanishes.
However, the above argument can still be employed in order to rationalize the variability of
energy gaps associated with different structural properties of the DB centre location. Small gaps single 
out highly connected regions, whose neighbourhood tends to be less tightly connected. 
By virtue of this correlation (also clearly illustrated by the plot of the energy gaps versus 
clustering coefficients; see Fig.~\ref{f:GAPS.cC3}), 
it is clear that the balance between the two above-mentioned mechanisms
is different for DBs centered at different sites. In particular, if the site is a highly 
connected one, the DB will harvest less energy in its tail, as compared to a DB 
localised in a poorly connected region, thus less effectively compensating 
the drop caused at its centre by the reduction in amplitude. Overall, this may yield a small energy gap. 
Conversely, a DB located in a loosely connected region will be 
more effective in counterbalancing the same reduction in energy by appealing to the contribution of
its more connected tail regions. Hence a higher energy gap.

The above argument provides a qualitative interpretation for the 
variability displayed by the energy gaps. Interestingly, 
when the structural properties of 
a large set of enzyme structures are investigated, 
such a rationalization also yields
a coherent, as well as intriguing, biological picture. 
Indeed, the cooling simulations reported in ref.~\cite{Juanico:2007}
showed very clearly that DBs tend to form spontaneously in the stiffest regions 
of a given enzyme, as identified using a simple indicator of local rigidity. 
Remarkably, catalytic sites, that is, 
the vital spots for the initiation of enzymatic activity,  
are also often found in such regions~\cite{Lavery:07,Juanico:2007}.
As an extension of such results,
the lower panels in Fig.~\ref{f:GAPS.cC3} report the distribution of 
connectivities and clustering coefficients for known catalytic sites 
in a set of 833 enzymes
from the 2.1.11 version of the catalytic site atlas~\cite{catal_atlas}. 
As a comparison, we calculated 
the same distributions for a random selection of residues of the same chemical type
from the same data set of structures.
Manifestly, catalytic sites tend to be highly connected and, 
accordingly, also show a marked tendency to have less inter-connected 
environments. In this sense, these conclusions fit within the picture drawn in ref.~\cite{Juanico:2007}, 
as there is an obvious positive correlation between degree of connectivity and stiffness, 
the more connected a region clearly the less easily deformable.

From the biological point of view, 
this observation suggests a straightforward interpretation 
for the association of small gaps with highly connected sub-domains.
Indeed, discrete breathers are more easily excitable in those regions where 
enzymes perform their activity.
Thus, the latter may use DBs in order
to lock down for relatively long periods of time
the energy released by ongoing chemical reactions, such as ATP hydrolysis.
While it is customary to appeal to energy storage mechanisms of chemical origin, 
we are here providing evidences that an additional {\em mechanical} channel, based on  
localised vibrations of non-linear origin, may exist. 
It is indeed tempting to imagine that the energy stored in a discrete breather could 
then be used to lower the barrier of a chemical reaction involved in a catalytic process. 
Because such a mechanism would drastically increase the efficiency 
of the enzyme~\footnote{Note that DB excitation over an energy gap would be a thermally-activated 
process in the presence of a thermal environment~\cite{Piazza:03}, and hence
even a small reduction factor in the threshold would magnify exponentially the excitation rate.}, 
it is likely that, if it is possible to implement it in an actual
protein structure and to make it work in a cellular environment, evolution
has found the way to do it. En passant, this may provide an answer to another,
long-standing question, namely: why are most enzymes so big, their active
site often occupying a single, tiny region of the whole structure ? 
Our results suggest that this might be so in order for them to have highly connected
parts where DBs can easily form and store high amounts of energy for relatively
long times, far enough away from the solvent and its dissipative effects.
 
%
\section{Conclusion and perspectives\label{sec:5}}
%

Using an analytical approach, we have corroborated our previous numerical 
results -- confirming, in particular, that the properties of discrete breathers in 
nonlinear network models of proteins are site-dependent. Moreover, 
we have shown that DBs of arbitrary low amplitude cannot be excited anywhere in 
the structure. For a few sites, namely, 
those associated to the largest displacements in the edge normal modes, the DB
energy goes to zero as its amplitude approaches zero. 
However, for the majority of sites a lower bound exists for the allowed DB amplitudes.
This, in turn, implies that the majority of sites host DBs that exist  but for 
energies higher than a certain site-dependent gap. Remarkably, we have shown 
non-zero gaps invariably arise as a consequence of the impossibility of exciting 
low-amplitude breathers in the first place.

While the profound origin of this puzzling phenomenology is still unclear, 
it is instructive to recall that similar phenomena arise in one-dimensional systems
in the presence of cubic plus quartic nonlinearities, essentially from the requirement of
strict convexness of the interaction potentials~\cite{aubry-k3-2001,James:2001}.
In our case, despite the fact that the interaction
potential between residues has the same functional form for all pairs, the equations 
of motion for a given residue in interaction with its neighbours 
contain quadratic, but also cubic nonlinearities, with coefficients that depend on the 
topology and on the spatial arrangement of its set of neighbours. Thus, it is tempting to speculate that 
spatial disorder and nonlinearity might team together, so as to produce a hierarchy of energy gaps
in the DB dispersion relations. Further work in this direction is currently under way.

Although interesting {\it per se},  that is, within the context of complex nonlinear networks,
our results may also prove to have a profound biological
significance. Indeed, we have shown that DBs can form more easily (small energy
gaps) in parts of the structure where connectivities are high and
clustering coefficients low. Reciprocally, we have shown that catalytic residues
tend to be highly connected and have low clustering coefficients.
However, in order to establish a link between 
these two facts on firmer grounds, it remains necessary to show that, 
like elastic network models, nonlinear network models of proteins are
able to capture, at the coarse-grained level, such key dynamical features of
actual protein structures. Work is in progress along these lines.

%
\section*{Glossary\label{sec:glos}}

\paragraph{\bf Discrete breathers}
Spatially localised, time-periodic solutions of the equations of motion arising
in many spatially extended discrete nonlinear systems. The frequency of the internal vibrational 
degree(s) of freedom lies  outside the linear spectrum (within gaps, if any) so that it does 
not resonate with higher harmonics of the linear modes. More loosely, the same denomination 
is employed to indicate generic long-lived, localised vibrations with nonlinear frequencies, 
as arising for example in periodic media from modulational instability of linear edge modes  
or from surface cooling. Despite the strong similarities, the question as to what is rigorously the 
link between the latter modes of vibrations (often also dubbed {\em chaotic} breathers)
and known exact solutions of the equations of motion remains open.

\paragraph{\bf Surface cooling}
Numerical method allowing for spontaneous localisation of energy in the form of chaotic breathers
in finite nonlinear discrete dynamical systems with free ends. Importantly, the method does not require any 
preliminary assumptions on the nature of the asymptotic localised modes.
In practice, the system is first thermalised and then cooled down by adding a friction term in the 
equations of motion of the particles sitting at the boundaries. In the case
of proteins, friction is put on solvent-accessible amino-acid residues.
Provided the initial energy density is above a given threshold (that may well be zero),
the system evolves toward a quasi-stationary state where all the residual energy of
the system is stored within a handful of sites and exponentially localised far from the boundaries. 
 
\paragraph{\bf Enzyme catalysis}
Enzymes are able to catalyze specific chemical reactions, that is,
to speed up their reaction by, typically, a factor of $10^9$. To do so, 
chemical reactants bind in a pocket of the structure, the enzyme active
site, where dedicated amino-acid residues, the so-called essential residues,
are used so as to lower the energy barriers involved in the chemical reaction.


\providecommand{\newblock}{}


\end{document}